# Radial Distribution of Production Rates, Loss Rates and Densities Corresponding to Ion Masses ≤40 amu in the Inner Coma of Comet Halley: Composition and Chemistry

S.A. Haider

Physical Research Laboratory, Ahmedabad 390009, India
haider@prl.ernet.in

and

Anil Bhardwaj[1]

Space Physics Laboratory, Vikram Sarabhai Space Centre, Trivandrum 695022, India
bhardwaj_spl@yahoo.com

[1] Currently NRC Senior Research Associate at the NASA Marshall Space Flight Center, NSSTC/XD12, 320 Sparkman Drive, Huntsville, AL 35805
Anil.Bhardwaj@msfc.nasa.gov

**Running Title: Ion Chemistry in Comet Halley's Ionosphere**

# ABSTRACT

In this paper we have studied the chemistry of C, H, N, O, and S compounds corresponding to ions of masses ≤40 amu in the inner coma of the comet 1P/Halley. The production rates, loss rates, and ion mass densities are calculated using the Analytical Yield Spectrum approach and solving coupled continuity equation controlled by the steady state photochemical equilibrium condition. The primary ionization sources in the model are solar EUV photons, photoelectrons, and auroral electrons of the solar wind origin. The chemical model couples ion-neutral, electron-neutral, photon-neutral and electron-ion reactions among ions, neutrals, electrons, and photons through over 600 chemical reactions. Of the 46 ions considered in the model the chemistry of 24 important ions (viz., $CH_3OH_2^+$, $H_3CO^+$, $NH_4^+$, $H_3S^+$, $H_2CN^+$, $H_2O^+$, $NH_3^+$, $CO^+$, $C_3H_3^+$, $OH^+$, $H_3O^+$, $CH_3OH^+$, $C_3H_4^+$, $C_2H_2^+$, $C_2H^+$, $HCO^+$, $S^+$, $CH_3^+$, $H_2S^+$, $O^+$, $C^+$, $CH_4^+$, $C_2^+$ and $O_2^+$) are discussed in this paper. At radial distances <1000 km, the electron density is mainly controlled by 6 ions, viz., $NH_4^+$, $H_3O^+$, $CH_3OH_2^+$, $H_3S^+$, $H_2CN^+$, and $H_2O^+$, in the decreasing order of their relative contribution. However, at distances >1000 km, the 6 major ions are $H_3O^+$, $CH_3OH_2^+$, $H_2O^+$, $H_3CO^+$, $C_2H_2^+$, and $NH_4^+$; along with ions $CO^+$, $OH^+$, and $HCO^+$, whose importance increases with further increase in the radial distance. It is found that at radial distances greater than ~1000 km (±500 km) the major chemical processes that govern the production and loss of several of the important ions in the inner coma are different from those that dominate at distances below this value. The importance of photoelectron impact ionization, and the relative contributions of solar EUV, and auroral and photoelectron ionization sources in the inner coma are clearly revealed by the present study. The calculated ion mass densities are compared with the Giotto Ion Mass Spectrometer (IMS) and Neutral Mass Spectrometer (NMS) data at radial distances 1500, 3500, and 6000 km. There is a reasonable agreement between the model calculation and the Giotto measurements. The nine major peaks in the IMS spectra between masses 10 and 40 amu are reproduced fairly well by the model within a factor of two inside the ionopause. We have presented simple formulae for calculating densities of the nine major ions, which contribute to the nine major peaks in the IMS spectra, throughout the inner coma that will be useful in estimating their densities without running the complex chemical models.

## 1. INTRODUCTION

The study of cometary composition has been the subject of great interest since 1986 when High Intensity Ion Mass Spectrometer (HIS-IMS), Neutral Mass Spectrometer (NMS), and Positive Ion Cluster Composition Analyzer of Reme Plasma Analyzer (PICCA-RPA) onboard Giotto spacecraft measured a large number of peaks between 12 and 120 amu in the mass spectra of comet 1P/Halley (Balsiger *et al*., 1986; Krankowsky *et al*., 1986; Mitchell *et al*., 1987). After the encounter of comet Halley, several theoretical models have been constructed describing the chemistry of cometary coma. But most of these models were limited to the chemistry of only few ions. Allen *et al*. (1987) derived the concentration of $NH_3$ and $CH_4$, relative to $H_2O$, by fitting the variation with distance of Giotto-IMS data. These authors showed that the measured ionosphere profiles were incompatible with pure $H_2O$ coma, and require 1-2% $NH_3$. The magneto-hydrodynamic-chemical model of Wegmann *et al*. (1987) (cf. also Schmidt *et al*., 1988) also yielded similar results. Marconi and Mendis (1988) reanalyzed the IMS data and found that an enhancement of solar EUV flux, responsible for ionization, can also account for the IMS observations.

Boice *et al*. (1990) suggested that most of the $CH_4$ estimated using the IMS data is not from volatiles released directly from the nucleus but is a by-product of polyoxymethylene (POM), a constituent of the complex organic mixture in CHON particle (Huebner *et al*., 1989). Boice et al. derived an upper limit of 0.5% for the ratio of $CH_4/CO$. However, Wegman *et al*. (1987) did include 3% POM in their model, and found that the data of IMS for singly charged ions with masses from 6 to 20 amu were in better agreement by including POM in the model than by excluding POM, but requires about 2% $CH_4$ in the nucleus. Ip *et al*. (1990) derived the relative abundance of HCN of 0.02% using a photochemical model and fitting the IMS data peak at 28 amu. Two sharp peaks observed in the PICCA data at 37 and 39 amu are identified as the organic ions $C_3H^+$ (Marconi et al., 1989) and $C_3H_3^+$ (Korth et al., 1989), respectively, whose sources are attributed to the circum-nuclear distribution of the CHON dust particles observed at the comet.

Geiss *et al*. (1991) reported that the IMS data between masses 25 and 35 amu are dominated by $H_3CO^+$ and $CH_3OH_2^+$, which give rise to peaks in the IMS data at 31 and 33 amu. Marconi *et al*. (1991) studied the chemistry of sulfur ions $H_3S^+$, $HCS^+$, $H_3CS^+$ and $SO^+$ corresponding to masses 35, 45, 47, and 48 amu, respectively, in the PICCA data. They have argued that $H_3S^+$ is a dominant ion in the chemistry of sulfur group ions, whose likely parent molecule is $H_2S$ (Marconi *et al*., 1990). Mitchell et al. (1992) extended the analyses of Giotto-PICCA data to higher masses. Meier *et al*. (1993) analyzed Giotto NMS data and found that the peak at 31 amu is dominated by protonated formaldehyde ion $H_3CO^+$, and that most of the $H_2CO$ comes from an extended source; the data can be reconciled even if $H_2CO$ is not a parent molecule that evaporates from the nucleus. Eberhardt *et al*. (1994) also suggested that the mass channels at 33 and 35 amu are dominated by methanol and hydrogen sulphide, respectively, and derived their relative abundances on the comet Halley.

Haider *et al*. (1993) solved coupled continuity equations for chemical steady state conditions using the chemistry of $H_2O$, $NH_3$ and $CH_4$. In this paper, the ratios of masses 19/18, 17/19 and 15/19 were determined and compared with Giotto IMS data to derive the abundances of $NH_3$ and $CH_4$ in the coma of comet Halley, which were about 1.5% $NH_3$ and 0.5% $CH_4$. Meier *et al*. (1994) also derived the abundance of ammonia using the NMS data and found results similar to those of Haider *et al*. (1993). Altwegg *et al*. (1994) analyzed the IMS-HIS data and reported that in the group of masses 12-16 amu the most abundant ion is $CH_3^+$ whose parent molecule could be $CH_2$. Haberli *et al*. (1995) developed a chemical-transport model to investigate the cause of sharp enhancement in ion density observed by Giotto IMS at a cometocentric distance of about 12000 km, by incorporating the chemistry of 5 major parent species in the model. Detailed analysis of the distribution of $H_2O^+$ ions in the coma of comet Halley has also been attempted recently using the MHD model with chemistry (Haberli *et al*., 1997; Wegmann *et al*., 1999). The study of Wegmann *et al*. (1999) suggested that about 3% of $H_2O$ is finally converted to $H_2O^+$ on comet Halley.

Bhardwaj *et al*. (1996) developed a coupled chemistry-transport model to study the chemistry of atomic carbon and oxygen in the inner coma of comet Halley. Their calculations revealed that the electron impact dissociation of parent species could be a potentially important source of C and O production and their emissions in the inner cometary coma. Later this calculation was expanded, and extended to investigate the chemistry of $C(^1D)$ atom and the mechanisms of generation of CI 1931 Å emission (Bhardwaj, 1999) in the inner coma of comet Halley, and the production of $O(^1D)$ atom and OI 6300 Å emission (Bhardwaj and Haider, 2002) on comet Wirtanen.

## 2. OBJECTIVE

The objectives of this paper are: 1) to study the chemistry of C, H, N, O, and S compounds corresponding to mass of ions ≤40 amu observed by HIS sensor of the Giotto IMS experiment in the inner coma (≤$10^4$ km) of comet Halley, and 2) to investigate how the different sources of ionization, viz., solar EUV photons, photoelectrons and auroral electrons of solar wind origin, play their role in different physical and chemical processes in the coma, and thereby affecting the production and density of these ions. The chemical modeling of all ionic species corresponding to 10–120 amu is very difficult because of rapidly increasing number of possible species at larger masses. Already more than 22 molecules have now been identified as parent molecules out-gassing from the nucleus (e.g., Crovisier, 1999). Moreover, several molecular and atomic species have been spectroscopically detected in the cometary coma at UV, visible, IR, and radio wavelengths. Although a few of these are simple parent molecules (i.e., $H_2O$, $CH_4$, CO, $CO_2$, $N_2$, HCN, $H_2CO$, $CH_3OH$, $NH_3$, $SO_2$, $H_2CS$, $CS_2$), most are radicals and ions that appear to be dissociation fragments of more complex, but unknown species. These molecules need not sublime directly from the nucleus. The challenge for the future modeling studies is to develop several new chemical reactions associated with the chemistry of ions >40 amu. It is possible that the organic species produced by radiation induced process in the cometary precursor grains within the solar nebula and/or in the interstellar medium could be the source of the compound at these masses.

In the present paper we have studied the chemistry of ions for masses ≤40 amu. The calculated densities are compared with the Giotto IMS and NMS data at all masses. Earlier Wegmann *et al*. (1987) carried out such study (cf. also Schmidt *et al*., 1988;

Huebner *et al*., 1991) that compared the model calculations with Giotto IMS data for ion masses up to 55 amu. However, they did not include the chemistry of $CH_3OH$ and $H_2S$ in their calculations. We have included these gases in our model. The methanol and $H_2S$ are important parent molecules that have been detected in several comets (cf. Eberhardt *et al*., 1994; Crovisier, 1999), and their protonated-ions, viz., $CH_3OH_2^+$ and $H_3S^+$, are among the major ionic species in the inner coma. Moreover, in the present study we have explicitly included the photoelectron and auroral electron impact ionization sources in the model and have evaluated their impact relative to photoionization source. Several authors have carried out calculations in order to fit the Giotto IMS data, but to the best of our knowledge, all of them, except Wegmann *et al*. (1987), presented their results only at few (≤5) ion masses. Several ionic species can contribute to a given ion mass and a detailed chemical model is required to get an insight into their relative contributions and to determine the dominant ion species. Also, it is important to investigate how model results with current chemistry compare with the Giotto IMS and NMS data under constraints of the abundances of parent species that have been detected and suggested by observations and modeling.

## 3. CALCULATION AND INPUT PARAMETERS

The present calculations are made at 1 astronomical unit for solar minimum condition appropriate for comet Halley's exploration period in 1986. Our chemical model couples ion-neutral, electron-neutral, photon-neutral, and electron-ion reactions among 46 ionic species, 15 neutral species, and electron through reactions numbering around 600. These reactions are compiled from a large number of references (prominent are Anicich and Huntress, 1986; Schmidt *et al*., 1988; Anicich, 1993a,b; Millar *et al*., 1997). The electron-ion recombination rates are taken from Mitchell (1990). The electron temperature is taken from Eberhardt and Krankowsky (1995) who derived it based on fitting ion population ratio of masses 19/18 and 33/19 with the Giotto NMS data. The neutral temperature is taken as 200 K. There will not be a significant change on the calculated values by changing the neutral temperature from 200 K to 100 K and 50 K (Bhardwaj, 1999).

Using the calculated production rates, the coupled equations were solved for chemical steady state condition for the densities of ions of masses ≤40 amu. The gases in the model atmosphere are taken 80% $H_2O$; and 1.5% $NH_3$, 0.5% $CH_4$, 3% $CO_2$, 0.1% $CS_2$, 0.1% $N_2$, 2% $C_2H_2$, 0.41% $H_2S$, 1.7% $CH_3OH$, 0.1% $SO_2$, and 0.1% HCN (all with respect to $H_2O$) which are constrained by various observations (cf. Festou *et al*., 1993; Haider *et al*., 1993; Eberhardt *et al*., 1994; Meier *et al*., 1994; Crovisier, 1999). The radial densities of these gases in the coma are calculated as described in Bhardwaj *et al*. (1995, 1996). The density distribution of CO and $H_2CO$ are taken from Eberhardt *et al*. (1987) and Meier *et al*. (1993), respectively, which include the extended source effects. The total gas production rate is taken as $1.3 \times 10^{30}$ s$^{-1}$ (Gringauz *et al*., 1986). The ion production rate due to photoionization and photodissociative ionization by solar UV-EUV radiation for the parent species are calculated following Bhardwaj *et al*. (1996) and Bhardwaj (1999, 2003), with rates for some of the parent species taken from Huebner *et al*. (1992). The photoionization rate for $SO_2$ is taken from Michael and Bhardwaj (1997).

These rates are calculated at solar zenith angle of 0°, but are valid for the entire sunlit hemisphere of the coma (cf. Bhardwaj, 2003), and are apt for the Giotto flyby of comet Halley whose closest approach occurred on the sunward side of the coma. A detailed discussion on the absorption of solar EUV radiation and degradation of EUV-generated photoelectrons in the inner coma, and more particularly close to the nucleus, is given in Bhardwaj (2003).

The photoelectron and auroral electron impact ionization and dissociative ionization rates are calculated using the Analytical Yield Spectrum approach (Bhardwaj *et al*., 1990, 1996; Bhardwaj, 1999, 2003, and references therein). The auroral electron impact production rates are calculated using solar wind electron spectrum measured by the plasma spectrometer (PLASMAG) experiment onboard VEGA 2 spacecraft on March 11, 1986 (Gringauz *et al*., 1986; Haider *et al*., 1993). Besides the VEGA-PLASMAG measurements, the electron energy spectra were also measured in the vicinity of Halley's comet by RPA-Copernicus experiment onboard Giotto (Reme *et al*., 1986). The PLASMAG observed a peak in the electron spectrum at energy ~1 keV (cf. Gringauz *et al*., 1986; Figure 3 in Haider *et al*., 1993), which was not seen in the RPA-Copernicus measurements at same distance from the nucleus. Mendis (1987) has argued that a "cometary aurora" was responsible for these energetic electrons, which were not detected during the Giotto flyby due to a very calm solar wind condition at that time. The cometary aurora occurs during the disturbed solar wind conditions due to partial or total disruption of cross-tail current system, which directs the current flow along the field lines that discharges through the head of the coma. The VEGA 2 spacecraft approached the coma of comet Halley during the disturb solar wind period when the electrons may be accelerated to a few keV due to cometary sub-storms events (cf. Bhardwaj *et al*., 1990, 1996; Haider *et al*. 1993; Bhardwaj, 1997, 1999).

In this paper the electron impact ionization and dissociative ionization cross sections for $H_2O$, $CH_4$, $CO$, $CO_2$, $N_2$, $CH_3OH$, $H_2CO$ and $NH_3$ are the same as those used by Bhardwaj *et al*. (1996) and Bhardwaj (1999). The electron impact cross sections for $CS_2$ are taken from Rao and Srivastava (1991), for $H_2S$ from Khare and Meath (1987), for $C_2H_2$ from Tawara *et al*. (1990) and for $SO_2$ from the compilation of Bhardwaj and Michael (1999). The ionization and dissociative ionization cross sections for HCN are not available. We have assumed these cross sections equal to that of $C_2H_2$ ionization and dissociative ionization cross sections, because HCN is isoelectronic as $C_2H_2$ and has quite similar photo-excitation cross sections (Haider and Bhardwaj, 1997).

## 4. CHEMISTRY

Table 1 shows the possible ion species that contribute to different peaks in the Giotto-IMS spectra from 12 to 40 amu. The ions for masses 22 amu and 23 amu are not shown. To account for ions having $^{13}C$, $^{18}O$, $^{33}S$, and $^{34}S$ corresponding to various ion masses, we have assumed solar distribution for isotopes; specifically $^{12}C : ^{13}C = 98.9 : 1.1$, $^{16}O : ^{18}O = 99.8 : 0.2$, and $^{32}S : ^{33}S : ^{34}S = 95.0 : 0.8 : 4.2$ in the model calculation. These values are consistent with those derived from Giotto observations (cf. Balsiger *et al*, 1995; Eberhardt *et al*., 1995; Altwegg *et al*., 1999). Our model calculates the densities of 46 ions ($CH_5^+$, $SH^+$, $S^+$, $H_2S^+$, $NH_3OH^+$, $CH_3^+$, $HCO^+$, $H_2CO^+$, $C_2H_3^+$, $C_3H_3^+$, $H_3CO^+$,

$H_2CN^+$, $H_3S^+$, $NH_3^+$, $CH_3OH_2^+$, $CH_3OH^+$, $NH_4^+$, $H_3O^+$, $N^+$, $NH^+$, $NH_2^+$, $N_2^+$, $C_2H^+$, $C^+$, $CH_2^+$, $CH^+$, $C_2N^+$, $C_3H_4^+$, $C_2H_4^+$, $C_2H_5^+$, $CO^+$, $O^+$, $H^+$, $C_2^+$, $HCN^+$, $OH^+$, $O_2^+$, $CHOH_2^+$, $HNO^+$, $N_2H^+$, $H_2O^+$, $CH_4^+$, $C_2H_2^+$, $(H_2O)_2^+$, $H_5O_2^+$, $C_3H^+$) in the inner coma of comet Halley.

Figure 1a shows the density profiles of nine important ions ($CH_3OH_2^+$, $H_3CO^+$, $NH_4^+$, $H_3S^+$, $H_2CN^+$, $H_2O^+$, $C_2H_2^+$, $C_3H_3^+$, and $H_3O^+$). The total ion density is also shown in the figure for with and without ("no-AE") auroral electron ionization source. The overall effect of auroral electron impact ionization source on the electron density is to increase its magnitude by ~50% at distances >100 km and to about a factor of two or more in the region very close to the nucleus. The IMS derived electron density along the Giotto inbound pass (Altwegg *et al*., 1993) is also shown in the Figure 1 for direct comparison of the model results. The sharp rise in the ion density at radial distance beyond 8000 km is due to rapid increase in the electron temperature at these radial distances (cf. Eberhardt and Krankowsky, 1995; Haberli *et al*., 1995) that results in a decrease in the electron-ion recombination loss rate.

Figure 1b shows the radial density profiles for 17 ions ($CH_3OH^+$, $C_3H_4^+$, $C_2H^+$, $HCO^+$, $S^+$, $CH_3^+$, $H_2S^+$, $O^+$, $C^+$, $CH_4^+$, $C_2^+$, $O_2^+$, $NH_3^+$, $OH^+$, $CO^+$, $H_5O_2^+$, and $(H_2O)_2^+$) that make a significant contribution to the various ion masses. The chemistry of all these 26 ions is discussed in this paper. The profiles for other ions are not shown because their densities are very low.

Figures 2a, 2b, and 2c show the model ion density spectra for masses between 12 and 40 amu at 1500 km, 3500 km, and 6000 km, respectively. These calculations are compared with the Giotto IMS data taken from Balsiger *et al*. (1986) at 1500 km and 6000 km, and from Altwegg *et al*. (1999) at 3500 km. Since Giotto IMS data is represented in arbitrary unit, it is normalized with the calculated ion density at mass 19 amu (i.e., the density of $H_3O^+$ ion, which is the dominant peak in the ion mass spectra). The general agreement between model and IMS measurement is good at 1500 km and 3500 km. In these figures, the measured peaks are reproduced well by the model within a factor of 2, except at mass 37 amu which is discussed later. In Figures 2a and 2b, we have also shown the ion densities measured by the Giotto NMS experiment at masses 17, 18, 19, 31, 33 and 35 amu. It should be noted that NMS data are given in absolute numbers and therefore no normalization is needed (Meier *et al*., 1993, 1994; Eberhardt *et al*., 1994). It is found that the model calculated densities are in good agreement with this measurement also. Thus, we can report that our model is able to explain consistently the Giotto IMS and NMS measurements inside the cometary ionopause.

At 6000 km, we find that though the calculated densities are in qualitative agreement with IMS data, they are consistently lower than the measured values. In particular, the model density is too low at 32 amu and is out of phase at 30 amu. One of the reasons for the discrepancy between the model and Giotto data at distances beyond the contact surface (~5000 km) could be the role played by the solar wind plasma through the magnetic field lines. Further, the assumption of photochemical equilibrium is not strictly valid at such distances, particularly for atomic ions since they have a longer lifetime. For example, $S^+$ is the main ion at mass channel 32 amu, which being an atomic ion has a much lower recombination rate and thus a longer lifetime. All the $S^+$ ions that are created on the sunward side outside the ionopause are carried with the solar wind back towards the contact surface: leading to a sharp increase in the density of $S^+$ near 6000 km (Altwegg *et al.*, 1999). Moreover, the charge exchange between solar wind plasma and cometary neutrals could also contribute to the coma chemistry outside the contact surface. None of these processes have been included in our coma chemistry model. Therefore, we suggest that our model results should be treated as approximate outside the cometary ionopause.

The calculated ion density at masses 13 and 14 amu is lower than the IMS measurement. The low abundance of these ions in the model could be due to not including $CH_2$ as a parent species in the cometary coma. Altwegg *et al.* (1994) have suggested that $CH_2$ should be present in the cometary nucleus, with relative abundance of about 0.25%, to get a close agreement between calculation and measurement at these masses. The calculated density at mass 37 amu is also lower than the observation (cf. Fig 2). This peak is produced mainly by $C_3H^+$ ion (Marconi *et al.*, 1989). The other possible source at this mass is $H_5O_2^+$ ion, which is produced from the chemistry of water dimer as given below (cf. Murad and Bochsler, 1987):

$(H_2O)_2 + h\nu \rightarrow (H_2O)_2^+ + e$   [K1=1.6x10$^{-7}$ s$^{-1}$]   (R1)
$(H_2O)_2^+ + H_2O \rightarrow H_5O_2^+ + OH$   [K2=1.0x10$^{-9}$ cm$^3$ s$^{-1}$]   (R2)

To assess the effects of including water dimer, we run the model by including dimer chemistry, as suggested by Murad and Bochsler (1987), varying its abundance from 0% to 5% relative to water. The results are shown in Figures 2a–2c (by open stars connected by dashed lines) for 1% dimer. The densities at masses 36 and 37 amu are found to increase sharply, but the densities at other ion masses are almost unaffected (and hence are not shown). The radial density profiles of $(H_2O)_2^+$ and $H_5O_2^+$ ions are presented in Fig. 1b for 1% dimer case. From Figures 2a–2c it appears that, by including about 1% dimer in the model, a reasonable agreement with the Giotto IMS data at all the three distances can be obtained. However, the relative abundance of water dimer being produced during the sublimation of ice at the cometary surface is dependent on the equilibrium temperature of ice on the nucleus, varying from about 0.3 at 200 K to 0.01 at 400 K (Slanina, 1986). With the surface temperature of comet Halley's nucleus suggested to be ~200-250 K, the amount of dimer formed would be 10-30%, which is too large a value than that required to consistently explain the Giotto IMS observations at 36 and 37 amu. Thus, it seems that, though dimer ions $(H_2O)_2^+$ and $H_5O_2^+$ makes an important contributions at ion masses 36 and 37 amu, the thermodynamics of water dimer formation seems inconsistent with the requirements of the Giotto IMS data.

**4.1. Chemistry of Ion Masses 12 to 15 amu**

In general there are 9 major peaks in Figures 2a-2c at masses 12, 15, 19, 26, 28, 31, 33, 35, and 39 amu. The first peak at 12 amu is contributed by $C^+$ ion only. The photon and electron impact dissociation of CO, followed by photodissociation of $CO_2$, are the major sources of $C^+$ production. The $C^+$ ion is essentially destroyed by $H_2O$ in the inner coma.

The ions responsible for mass 13 amu are $CH^+$ and $^{13}C^+$, and that for the 14 amu are $N^+$, $CH_2^+$, and $^{13}CH^+$. The second peak, at 15 amu, is produced mainly from the $CH_3^+$ ion. This ion does not react with $H_2O$. Deep (<1000 km) in the coma, the $CH_3^+$ ion is destroyed mainly in reaction with parent species $CH_3OH$, $C_2H_2$, and $NH_3$. The dissociative recombination of $CH_3^+$ is the major loss process at larger distances (>1000 km) from the nucleus. The $CH^+$ and $CH_2^+$ ions react strongly (K ~$1.0 \times 10^{-9}$ cm$^3$ s$^{-1}$) with most of the parent neutral species and produce several major and minor ions in the cometary coma. The density of $NH^+$ is very low as compared to that of the $CH_3^+$, and therefore $NH^+$ does not contribute much in the formation of the peak at 15 amu. The $N^+$ and $NH^+$ ions are produced by dissociation (photon and electron impact) of $N_2$ and $NH_3$, respectively, and are mainly destroyed in reaction with $H_2O$. The relative contributions of the major production and loss channels of these ion masses are given in Tables 2 and 3.

### 4.2. Chemistry of Ion Masses 16 to 18 amu

The ions responsible for the mass at 16 amu are $NH_2^+$, $O^+$, $^{13}CH_3^+$, and $CH_4^+$. The electron and photon impact dissociative ionization of $H_2O$, CO and $CO_2$ are the major sources of $O^+$ ion, which is mainly destroyed by $H_2O$ throughout the inner coma. The dissociation of $NH_3$ is the main source of $NH^+$ and $NH_2^+$, which are entirely destroyed by $H_2O$. The parent species $CH_4$, $NH_3$, $H_2O$, $H_2CO$, $H_2S$, and $CH_3OH$ also take part in the destruction of $NH_2^+$. But their contribution is less than 1% in the Cometary coma. The ionization of methane by photons, photoelectrons and auroral electrons are the main sources of $CH_4^+$ ion. This ion reacts with all major parent species, but is destroyed mainly by $H_2O$ with small contribution through dissociative recombination process (cf. Table 2).

The ions at mass 17 amu are taken as $OH^+$, $CH_5^+$, $^{13}CH_4^+$ and $NH_3^+$. The impact of photon, photoelectron and auroral electrons is found to be important mechanism for the production of $OH^+$ ion. The ions $CH_4^+$, $OH^+$ and $CH_5^+$ are predominantly destroyed by $H_2O$ resulting in the formation of $H_3O^+$ and $H_2O^+$ ions. The other parent species $C_2H_2$, $NH_3$, O, CO, $CO_2$, OH, H, C, and $H_2CO$ also participate in the sink process of $CH_5^+$. However, their percentage contributions are significantly low. Close to the nucleus, the $NH_3^+$ ion is mainly destroyed with $H_2O$. Above 1000 km dissociative recombination is the major loss process of $NH_3^+$.

The channel 18 contains $H_2O^+$ and $NH_4^+$ ions. The reaction of $H_3O^+$ with $NH_3$ is the most important source of this ion, which is destroyed only by dissociative recombination reaction. Addition of a small amount of $NH_3$ opens a channel for transferring $H_3O^+$ ions to $NH_4^+$, thereby increasing the density of $NH_4^+$ ion making it the dominant ion at distances close (<200 km) to the nucleus (cf. Fig. 1a). The density of $NH_4^+$ in the inner coma can be approximately calculated by the equation

$$[NH_4^+] = \frac{[H_3O^+][NH_3] K3}{[Ne] K4] \tag{A1}$$

where K3=$2.2 \times 10^{-9}$ cm$^3$ s$^{-1}$, K4=$4.15 \times 10^{-5}$ $(1/Te)^{0.5}$ cm$^3$ s$^{-1}$, and Ne is the electron density. Fig. 1a also shows the density profile of $NH_4^+$ ion (shown by open circles, which closely follow the solid line profile marked as $NH_4^+$) obtained by using equation (A1), demonstrating that values obtained from (A1) are a good representative of those obtained by running the extensive chemical model.

At channel 18 amu, $NH_4^+$ is the major ion at distances <800 km, while at larger distances (>1500 km) the $H_2O^+$ ion dominates. Photoionization, photoelectron and auroral electron impact ionization are the dominant sources of $H_2O^+$, but below 50 km the auroral electron impact ionization is the major source of $H_2O^+$. The major loss of $H_2O^+$ at all radial distances is its reaction with its parent species $H_2O$. About 25-30% $H_2O^+$ is produced from the reactions between $H_2O$ and $CO^+$, $OH^+$, and $H^+$ ions. The density of $H_2O^+$ ion in the inner coma can be calculated approximately as given below:

$$[H_2O^+] = \frac{[H_2O]\{h\nu + PE + AE\} + [H_2O]\{[CO^+]K6 + [OH^+]K7 + [H^+]K8\}}{[H_2O]K5} \quad (A2)$$

where $h\nu$, PE, and AE represent solar EUV photon, photoelectron and auroral electron sources, respectively, K5=1.85x10$^{-9}$ cm$^3$ s$^{-1}$, K6=1.9x10$^{-9}$ cm$^3$ s$^{-1}$, K7=1.6x10$^{-9}$ cm$^3$ s$^{-1}$, and K8=8.2x10$^{-9}$ cm$^3$ s$^{-1}$. The density profile of $H_2O^+$ ion obtained by using equation (A2) is shown in Fig. 1a (open circles, which closely follow the solid line profile marked as $H_2O^+$) signifying that the values obtained from (A2) well represent those obtained by running the extensive chemical model. Nearly 2-6 % loss of $H_2O^+$ is contributed from CO and $NH_3$. The loss of $H_2O^+$ caused by other parent species $CH_4$, $H_2S$, HCN and $SO_2$ are found to be less than 1%. The electron dissociative recombination loss of $H_2O^+$ is the second important loss mechanism at distance > 2000 km. The reaction of $H_2O$ with $N^+$, $N_2^+$, $NH^+$, $HCN^+$ and $C^+$ are significant source of $H_2O^+$ ions. The percentage contribution of major production and loss reactions for ion masses 16 to 18 amu are given in Tables 2 to 4.

### 4.3. Chemistry of Ion Masses 19 to 21 amu

The third peak in Fig. 2 is at mass 19 amu, which is the dominant peak among all the peaks, and is due to the $H_3O^+$ ion. The loss of $H_2O^+$ with its parent species is the main source of $H_3O^+$ ion. At shorter distances (<300 km), one of the major sink of $H_3O^+$ is its proton-transfer reaction with $CH_3OH$, which in fact is the dominant source of the $CH_3OH_2^+$ ion. The dissociative recombination of $H_3O^+$ plays the role of a major loss mechanism at distances >300 km from the nucleus. Nearly 5-15 % $H_3O^+$ is produced due to losses of $HCO^+$ and $OH^+$ with $H_2O^+$. Closer to the nucleus (<100 km) about 5-10 % $H_3O^+$ ions are destroyed through reaction with $H_2S$ and HCN. The production of $H_3O^+$ due to the loss of ions $H_2S^+$, $CH_4^+$, $NH_2^+$, $H_2CO^+$, $C_2H_3^+$, $SH^+$, and $HCN^+$ with $H_2O^+$ is found to be insignificant. These reactions contribute < 0.1 % $H_3O^+$ ion. The loss of $H_2O^+$ with $CH_4$ is also very small (0.1%) source of $H_3O^+$ ion. The reactions due to $H_3CO^+$-$H_2O$ and $H_3O^+$-$H_2CO$ contribute <10 % production and loss of $H_3O^+$ respectively. Neglecting the minor sources in the chemistry of $H_3O^+$ ion in the inner coma, the density of $H_3O^+$ ion can be approximately calculated as follows:

$$[H_3O^+] = \frac{[H_2O^+][H_2O]K5}{[CH_3OH]K9 + [NH_3]K3 + [N_e]K10} \quad (A3)$$

where K9=2.5x10$^{-9}$ cm$^3$ s$^{-1}$ and K10=2.33x10$^{-7}$ (300/Te)$^{0.5}$ cm$^3$ s$^{-1}$. The density profile of $H_3O^+$ ion obtained by using equation (A3) is depicted in Fig. 1a (line with open circles, which closely follow the solid line profile marked as $H_3O^+$).

The densities of ions $H_2{}^{18}O^+$ and $H_3{}^{18}O^+$, corresponding to masses 20 amu and 21 amu, respectively, are calculated by taking oxygen isotope $^{18}O$ as 0.2% of $^{16}O$ in the chemical model. The important production and loss channels of $H_3O^+$ and their relative contributions are shown in Table 4.

### 4.4. Chemistry of Ion Masses 24 to 26 amu

The channels 24 amu and 25 amu are contributed by $C_2{}^+$ and $C_2H^+$ ions, respectively. The ion $C_2{}^+$ is produced mainly in electron impact dissociation of $C_2H_2$, and is lost by dissociative recombination process and in reaction with $C_2H_2$. The dissociative ionization of $C_2H_2$ by photon, photoelectron, and auroral electron are the major sources of $C_2H^+$ ion, which is essentially destroyed by parent species $C_2H_2$ at lower (<500 km) radial distances and by dissociative recombination at larger distances. Table 5 quantify the relative contribution of the major production and loss reactions of $C_2{}^+$ and $C_2H^+$ ions.

In Figure 2, the fourth peak is produced at mass 26 amu corresponding to the ion $C_2H_2{}^+$. The production and loss rates of this ion are presented in Figure 3. The dominant production processes of this ion are photoionization, and photoelectron and auroral electron impact ionization of $C_2H_2$. It should be noted that, even in the absence of auroral ionization source, the photoelectron impact ionization source is equally important as the photoionization source in the formation of $C_2H_2{}^+$ ion, and in fact the former dominates at distances <200 km. The $C_2H_2{}^+$ ion is lost in the formation of major ions $H_3O^+$, $NH_4{}^+$, and $CH_3OH_2{}^+$ in the cometary coma via reactions with $H_2O$, $NH_3$, and $CH_3OH$, respectively. The dissociative recombination is the major loss process at radial distances beyond 3000 km. Nearly, 10 % $C_2H_2{}^+$ ion is destroyed by $C_2H_2$, $NH_3$ and $CH_3OH$. The other loss processes of $C_2H_2{}^+$ are found insignificant by its charge transfer reactions with $H_2S$ and $CH_4$. Nearly 1% $C_2H_2{}^+$ ions are produced due to loss of $CO^+$ and $C_2H^+$ with $C_2H_2$ and $CH_4$, respectively. Very small amount (<1%) of the ion is created by reaction $CH_4{}^+ + C_2H_2$. Neglecting the minor sources in the chemistry of $C_2H_2{}^+$ ion in the inner coma, the density of $C_2H_2{}^+$ ion can be approximately calculated as follows:

$$[C_2H_2{}^+] = \frac{[C_2H_2]\,[h\nu + PE + AE]}{[H_2O]K11+[NH_3]K12+[Ne]K13+[CH_3OH]K14+[C_2H_2]K15} \quad (A4)$$

where K11=$2.2\times10^{-10}$ cm$^3$ s$^{-1}$, K12=$9.6\times10^{-10}$ cm$^3$ s$^{-1}$, K13=$6\times10^{-7}$ $(300/Te)^{0.5}$ cm$^3$ s$^{-1}$, K14=$3.4\times10^{-10}$ cm$^3$ s$^{-1}$, and K15=$1.39\times10^{-9}$ cm$^3$ s$^{-1}$. The density profile of $C_2H_2{}^+$ ion obtained by using equation (A4) is presented in Fig. 1a (line with open circles, which closely follow the solid line profile marked as $C_2H_2{}^+$).

### 4.5. Chemistry of Ion Masses 27 to 29 amu

The mass channel 27 contains $HCN^+$ and $C_2H_3{}^+$ ion. Ionization of HCN by photons and electrons are the dominant sources of $HCN^+$ ion. The charge transfer reaction of $CO^+$ with HCN is also important source contributing about 10-15% of $HCN^+$. This ion is dominantly destroyed with $H_2O$. The loss by dissociative recombination of $HCN^+$ is important only at >3000 km, but is always less than the former in the inner coma. The ion $C_2H_3{}^+$ is mainly produced by the loss of $HCO^+$ and $C_2H^+$ with $C_2H_2$. The dominant loss of this ion occurs with $H_2O$ resulting in the production of the major ion $H_3O^+$. The dissociative recombination is second major loss mechanism only above 2000 km. The loss reactions of $C_3H^+$ with $H_2O$, $C_2H_2^+$ with $CH_4$ and $CH_4^+$ with $C_2H_2$ also facilitate in the formation of $C_2H_3^+$ ion totaling to about 10-15 %.

There are four ions, namely, $N_2^+$, $C_2H_4^+$, $CO^+$ and $H_2CN^+$, for the fifth peak at 28 amu in the ion mass spectra (cf. Figure 2). The peak at this channel is dominantly produced from ions $H_2CN^+$ and $CO^+$, since the density of ions $N_2^+$ and $C_2H_4^+$ are relatively very small. The ion $N_2^+$ is produced by photoionization, and photoelectron and auroral electron impact ionization of $N_2$ molecule. This ion is lost significantly by all parent species through charge exchange and dissociative recombination reactions. The loss of $CH_2^+$ with $CH_4$ is the dominant source of $C_2H_4^+$ ion, which is destroyed by dissociative recombination and also with $NH_3$ and $C_2H_2$.

The major production and loss channels of $CO^+$ ion and their relative contributions are presented in Table 5. This ion is mainly produced by ionization of CO, while major sinks of this ion is charge exchange between $CO^+$ and $H_2O$ at distances <300 km and reaction with $H_2CO$ at >1000 km. The other sink reactions of $CO^+$ caused by $CH_3OH^+$, $NH_3$, $CO_2$, $H_2S$, $CH_4$, $C_2H_2$ and HCN contribute less than 1% in the loss chemistry of $CO^+$ ion. Therefore, the percentage contributions due to these reactions are not given in Table 5. The production and loss rates of $H_2CN^+$ are shown in Figure 4. The loss of ion $H_3O^+$ with HCN is the main source of $H_2CN^+$. The other significant source of $H_2CN^+$ is obtained due to the loss of $H_3S^+$ ion with HCN. The main sink process of $H_2CN^+$ below 600 km is its reaction with $NH_3$ producing dominant ion $NH_4^+$. The electron recombination process is the dominant loss mechanism above 2000 km. The $H_2CN^+$ reacts very weakly (K=8.8x10$^{-13}$ cm$^3$ s$^{-1}$) with dominant parent species $H_2O$ producing $H_3O^+$, and hence $H_2CN^+$ densities are significantly large (cf. Fig. 1a) even for a very small abundance of HCN in the coma. The loss of this ion due to $H_2CO$ is very small near nucleus but is significant at larger distance, while for $H_2S$ it is other way. Very small amount of $H_2CN^+$ ion is produced due to loss of $H_3CO^+$ and $H_2O^+$ with HCN. Neglecting the minor sources of production and loss of the $H_2CN^+$ ion, its density in the inner coma can be approximately calculated by the equation:

$$[H_2CN^+] = \frac{0.5\,[H_3O^+][HCN]K17 + [H_3S^+][HCN]K18}{[NH_3]K16 + [Ne]K19} \quad (A5)$$

where K16=2.4x10$^{-9}$ cm$^3$ s$^{-1}$, K17=3.8x10$^{-9}$ cm$^3$ s$^{-1}$, K18=1.9x10$^{-9}$ cm$^3$ s$^{-1}$, and K19=2.13x10$^{-7}$ (300/Te)$^{0.5}$ cm$^3$ s$^{-1}$.

The channel 29 consists of $N_2H^+$, $HCO^+$, and $C_2H_5^+$ ions. The ion $N_2H^+$ reacts strongly (K ~1 to 4 x 10$^{-9}$ cm$^3$ s$^{-1}$) with all major parent molecular species in the cometary coma, and is destroyed dominantly by water because of its highest abundance. This ion is mainly produced in reactions involving either $N_2$ or its ion $N_2^+$. However, the chemistry of this ion does not play a significant role since the density of this ion is too low. The ion $C_2H_5^+$ is produced by positive ion-atom interchange reaction of $CH_3^+$ and $CH_2^+$ with methane, and is destroyed mainly by its efficient interaction (K=1.9x10$^{-9}$ cm$^3$ s$^{-1}$) with water forming major ionic species $H_3O^+$. The major ion at mass 29 amu is $HCO^+$. The major production and loss processes of ion $HCO^+$ are quantified in Table 5. This ion is mainly destroyed in reaction with $H_2O$ that results in the formation of $H_3O^+$ ion. The dominant production for this ion is obtained from the destruction of $H_2O^+$ and $CO^+$ with CO and $H_2O$ respectively. The loss of $C^+$ with $H_2O$ produces nearly 10% $HCO^+$ ion in the coma. The dissociative recombination process contribute <5% in the loss process of $HCO^+$ ion.

**4.6. Chemistry of Ion Masses 30 to 33 amu**

The ions contributing to mass 30 amu are $H_2CO^+$ and $CH_2NH_2^+$, whose densities are very small in the inner coma of comet Halley. Several parent species take part in the formation and destruction of the ion $H_2CO^+$. Among them photoionization and electron impact direct ionization of $H_2CO$, dissociative ionization of $CH_3OH$, and charge-transfer reaction of $H_2O^+$ with $H_2CO$, are the dominant production processes of $H_2CO^+$ ion. The main loss of this ion occurs in reaction with $H_2O$. The losses of ions $CH_2^+$ and $CH_3^+$ in reaction with $NH_3$ are the source reactions of $CH_2NH_2^+$ ion, which is mainly destroyed by dissociative recombination process.

The sixth peak in Figure 2, at mass 31 amu, is formed from ions $H_3CO^+$ and $HNO^+$. The density of $HNO^+$ is very low compared to $H_3CO^+$, and therefore $HNO^+$ is not an important ion for mass channel 31 amu. In Figure 5 we present the radial profiles of important formation and destruction processes of $H_3CO^+$ ion. The maximum loss of $H_3CO^+$ occurs in reaction with $H_2O$ that results in the formation of $H_3O^+$. However, at radial distances >1000 km, dissociative recombination is also an equally important mechanism which in fact leads at distances >5000 km. The formation of ammonium ion is not significantly large due to loss of $H_3CO^+$ with $NH_3$. Below 50 km, the $H_3CO^+$ ion is produced mainly due to charge transfer reactions between methanol and $CO^+$ and $CH_3^+$ ions, while at distances >100 km the main production source of $H_3CO^+$ is obtained through loss of $H_3O^+$ with $H_2CO$. Deep in the coma about 10% $H_3CO^+$ ions are produced through the destruction of $CH_2^+$ with $H_2O$ and of $O^+$ with $CH_3OH$. Neglecting the minor sources of production and loss of the $H_3CO^+$ ion, its density in the inner coma can be calculated by the equation:

$$[H_3CO^+] = \frac{[H_3O^+][H_2CO]K20 + [CO^+][CH_3OH]K21 + [CH_3^+][CH_3OH]K22 + [H_2O^+][H_2CO]K23}{[H_2O]K24 + [Ne]K25 + [NH3]K26}$$

(A6)

where K20=$3.6 \times 10^{-10}$ cm$^3$ s$^{-1}$, K21=$2.4 \times 10^{-9}$ cm$^3$ s$^{-1}$, K22=$2.3 \times 10^{-9}$ cm$^3$ s$^{-1}$, K23=$6.6 \times 10^{-10}$ cm$^3$ s$^{-1}$, K24=$2.3 \times 10^{-10}$ cm$^3$ s$^{-1}$, K25=$5.0 \times 10^{-7}$ (300/Te)$^{0.5}$ cm$^3$ s$^{-1}$, and K26=$1.7 \times 10^{-9}$ cm$^3$ s$^{-1}$. In Fig. 1a we show the density profile of $H_3CO^+$ ion obtained by using equation (A6) (line with open circles, which closely follow the solid line profile marked as $H_3CO^+$), which demonstrates that the values obtained from (A6) are a reasonably good representative of those obtained by running the extensive chemical model.

The channel at mass 33 amu shows a well defined peak at 1500 and 3500 km (as seen from Figures 2a and 2b), but at 6000 km (cf. Figure 2c) channels for 32 amu and 33 amu produce a broad maxima in the ion density spectrum. The ions responsible for these channels are $O_2^+$, $S^+$, $CH_3OH^+$, $SH^+$ and $CH_3OH_2^+$. The density of $O_2^+$ is very low relative to $S^+$ and $CH_3OH^+$ (cf. Figure 1b) and can be neglected in the formation of peak at mass 32 amu. Thus, the major ions for this channel are $S^+$ and $CH_3OH^+$. The $S^+$ ion is produced mainly by dissociative ionization of $H_2S$, $CS_2$, and $SO_2$ by solar EUV photon, auroral and photoelectrons, and is dominantly lost in reaction with $NH_3$ and $C_2H_2$ (cf. Table 6). The major loss of $S^+$ occurs with $C_2H_2$ and $NH_3$. The loss of this ion due to $H_2CO$ is significant (~10%) above 1000 km. The main production sources of $CH_3OH^+$ ion are photon, photoelectron, and auroral electron impact ionization of $CH_3OH$. This ion is mainly destroyed with $H_2O$ and electron (cf. Table 5).

The density of $SH^+$, corresponding to ion mass 33 amu, is negligible compared to that of the $CH_3OH_2^+$ ion. The important production and loss rates of the major ion $CH_3OH_2^+$ are represented in Figure 6. The destruction of $H_3O^+$ with methanol is the main source of ion $CH_3OH_2^+$. This ion is destroyed mainly by dissociative recombination at >~1000 km, while at lower distances it is lost in charge transfer processes with $NH_3$ and $CH_3OH$ molecules. The sinks of ions $HCO^+$, $C_2H_2^+$, $C_3H_4^+$, $NH_2^+$ and $H_3CO^+$ with methanol are minor reactions that combined together producing about <5% $CH_3OH_2^+$. The loss rate of this ion with $H_2CO$ is roughly equal to the production rate obtained from the destruction of $H_3CO^+$ with $CH_3OH$. Thus, ion density of $CH_3OH_2^+$ in the inner coma can be calculated using:

$$[CH_3OH_2^+] = \frac{[H_3O^+][CH_3OH]K9}{[NH_3]K27 + [CH_3OH]K28 + [Ne]K29} \quad (A7)$$

where K27=8.0x10$^{-7}$ (300/Te)$^{0.5}$ cm$^3$ s$^{-1}$, K28=2.0x10$^{-9}$ cm$^3$ s$^{-1}$, and K29=7.6x10$^{-10}$ cm$^3$ s$^{-1}$. Fig. 1a shows the density profile of $CH_3OH_2^+$ ion obtained by using equation (A7) (line with open circles), which almost overlap the solid line profile marked as $CH_3OH_2^+$, revealing that values obtained from (A7) are almost the same as those obtained by running the extensive chemical model.

**4.7. Chemistry of Ion Masses 34 to 36 amu**

The channel 34 amu consists of two ions, viz., $H_2S^+$ and $^{34}S^+$. The major sources of the formation of $H_2S^+$ are ionization of $H_2S$ by solar EUV, photoelectron, and auroral electron, and proton-transfer reactions of $H_2O^+$, $C_2H_2^+$, and $CO^+$ ions with $H_2S$ (cf. Table 6). The $H_2S^+$ ion is lost mainly in reaction with $H_2O$ forming the $H_3O^+$ ion; but, above 1000 km dissociative recombination process also makes a significant contribution and its importance increases with increasing radial distance (cf. Table 6). The other minor productions of this ion are obtained through destruction of $OH^+$ and $O^+$ with $H_2S$. The percentages of these reactions are not given in Table 6 because of their insignificant contribution in the coma.

In Figure 2, the eighth peak at 35 amu corresponds to ion $H_3S^+$ whose production and loss rates are shown in Figure 7. The most important source of $H_3S^+$ is obtained due to loss of $H_3O^+$ with $H_2S$. The major sink of this ion occurs due to its collision with $NH_3$, while at larger (>1000 km) distances dissociative recombination is the major loss mechanism. Charge exchange processes with water and hydrogen cyanide also destroy this ion, amounting to about 15% at <100 km. Nearly 5 % $H_2S^+$ ion is produced by loss of $H_2O^+$ and $H_2CN^+$ with $H_2S$. The destruction of ions $OH^+$, $HCO^+$ and $H_3CO^+$ with $H_2S$ are the minor sources of $H_3S^+$. The major sink of this ion occurs due to its collision with $NH_3$

, which produces dominant ion $NH_4^+$. The density of $H_3S^+$ ion in the inner coma can be calculated approximately from following equation:

$$[H_3S^+] = \frac{[H_3O^+][H_2S]K30}{[NH_3]K32 + [Ne]K31} \quad (A8)$$

where K30=1.4x10$^{-9}$ cm$^3$ s$^{-1}$, K31=3.7x10$^{-7}$ (300/Te)$^{0.5}$ cm$^3$ s$^{-1}$, and K32=1.9x10$^{-9}$ cm$^3$ s$^{-1}$. Fig. 1a presents the density profile of $H_3S^+$ ion obtained by using equation (A8) (line with open circles, which overlaps the solid line profile marked as $H_3S^+$).

There are two isotopes, viz., $H_2^{34}S^+$ and $H_3^{33}S^+$ that are taken into account to form ion channel of mass 36 amu. About 2.5% densities are produced from these compounds as compare to that obtained from equation (A8) for the ion $H_3S^+$.

## 4.8. Chemistry of Ion Masses 37 to 40 amu

The ions $C_3H^+$, $C_2N^+$, and $C_3H_3^+$ are taken in the spectra corresponding to masses 37 amu, 38 amu and 39 amu, respectively. The densities of ions $C_3H^+$ and $C_2N^+$ are nearly equal to the ion of mass 13 amu. The ion $C_3H^+$ is dominantly produced by acetylene. The major loss of this ion occurs with water, methanol and electron. The loss of $C^+$ with HCN produces $C_2N^+$ ions, which is destroyed mainly with $H_2O$. These reactions do not play a significant role in the chemistry of cometary coma. In Figure 2, the ninth peak is obtained from production and loss processes of the ion $C_3H_3^+$. The rates of the important processes for the $C_3H_3^+$ ion are presented in Figure 10. The main loss of this ion at shorter (<1000 km) radial distances occurs due to its chemical reaction with $C_2H_2$ (K=1.0x10$^{-9}$ cm$^3$ s$^{-1}$), while at larger distances dissociative recombination loss dominates. The $C_3H_3^+$ ion is mainly produced by the reaction of $CH_3^+$ with $C_2H_2$ (K=1.2x10$^{-9}$ cm$^3$ s$^{-1}$). Nearly 10% of $C_3H_3^+$ ion is formed in reactions of $C_2H^+$ and $C_2H_2^+$ ions with $CH_4$. The small amounts of $C_3H_3^+$ ions are produced through losses of $CH_4^+$ and $CH_2^+$ with $C_2H_2$ and $C_3H^+$ with $CH_4$. Neglecting the minor sources of production and loss of the $C_3H_3^+$ ion, its density in the inner coma can be calculated by using three reactions as follows:

$$[C_3H_3^+] = \frac{[CH_3^+][C_2H_2]K35}{[C_2H_2]K33 + [Ne]K34} \quad (A9)$$

where K33=1.0x10$^{-9}$ cm$^3$ s$^{-1}$, K34=1.0x10$^{-7}$ (300/Te)$^{0.5}$ cm$^3$ s$^{-1}$, and K35=1.2x10$^{-9}$ cm$^3$ s$^{-1}$. Fig. 1a shows the density profile of $C_3H_3^+$ ion obtained by using equation (A9) (line with open circles), which closely follows the solid line profile marked as $C_3H_3^+$.

The ions corresponding to 40 amu are $CH_2CN^+$ and $C_3H_4^+$. The $CH_2CN^+$ is obtained due to the loss of ion $CH_2^+$ with HCN and is mainly destroyed by dissociative recombination process. The $C_3H_4^+$ ions are produced in reactions of $C_2H^+$ and $C_2H_2^+$ with $CH_4$. The $C_3H_4^+$ ion is mainly lost in reaction with $C_2H_2$ and by dissociative recombination process, where the former and latter processes are dominating at altitudes <100 km and >500 km respectively (cf. Table 5).

## 5. SUMMARY AND DISCUSSION

In the present paper we have developed a chemical model to study the chemistry of 46 ions ($CH_5^+$, $SH^+$, $S^+$, $H_2S^+$, $NH_3OH^+$, $CH_3^+$, $HCO^+$, $H_2CO^+$, $C_2H_3^+$, $C_3H_3^+$, $H_3CO^+$, $H_2CN^+$, $H_3S^+$, $NH_3^+$, $CH_3OH_2^+$, $CH_3OH^+$, $NH_4^+$, $H_3O^+$, $N^+$, $NH^+$, $NH_2^+$, $N_2^+$, $C_2H^+$, $C^+$, $CH_2^+$, $CH^+$, $C_2N^+$, $C_3H^+$, $C_2H_4^+$, $C_2H_5^+$, $CO^+$, $O^+$, $H^+$, $C_2^+$, $HCN^+$, $OH^+$, $O_2^+$, $CHOH_2^+$, $HNO^+$, $N_2H^+$, $H_2O^+$, $CH_4^+$, $C_2H_2^+$, $(H_2O)_2^+$, $H_5O_2^+$ and $C_3H^+$) corresponding to masses ≤40 amu in the inner coma of comet Halley. The ionization sources included in the model are solar EUV photons, photoelectrons, and auroral electrons of solar wind origin. The production rates, loss rates, and mass densities of these ions are calculated using the Analytical Yield Spectrum approach and coupled continuity equation controlled by steady state photochemical model, which involves over 600 chemical reactions among ions, neutrals, photons, and electrons in the coma.

There are 24 important ions, viz., $CH_3OH_2^+$, $H_3O^+$, $NH_4^+$, $H_3S^+$, $H_2CN^+$, $H_2O^+$, $CO^+$, $C_2H_2^+$, $HCO^+$, $OH^+$, $NH_3^+$, $H_3O^+$, $CH_3OH^+$, $C_3H_4^+$, $C_2H^+$, $S^+$, $C_3H_3^+$, $CH_3^+$, $H_2S^+$, $O^+$, $C^+$, $CH_4^+$, $C_2^+$, and $O_2^+$, whose chemistry is discussed in this paper. At shorter (<1000 km) radial distances, the electron density is found to be overwhelmingly controlled by 6 ions, viz., $NH_4^+$, $H_3O^+$, $CH_3OH_2^+$, $H_3S^+$, $H_2CN^+$, and $H_2O^+$. The coma of Halley's comet is optically thin and the attenuation of solar EUV flux takes place within few 100 km from the nucleus. In the case of bright/dusty comets like Hale Bopp, the atmosphere near nucleus becomes optically thick and the photo ionization peak extends to few 1000 km from the nucleus (Bhardwaj, 2003). At larger distances (>1000 km), the 6 major ions are $H_3O^+$, $CH_3OH_2^+$, $H_2O^+$, $H_3CO^+$, $C_2H_2^+$, and $NH_4^+$; along with ions $CO^+$, $OH^+$, and $HCO^+$, whose importance increases with increase in the radial distance. Figure 9 shows the schematic diagram for the chemistry of important reactions channels, which are described in this paper.

In the formation of inner cometary coma, the contributions of important ions near the peak density of most dominant ions (~300 km) are nearly 40% $H_3O^+$, 26% $NH_4^+$, 20% $CH_3OH_2^+$, 5.3% $H_3S^+$, 2.66% $H_2CN^+$, 1.66% $H_2O^+$, 0.35% $NH_3^+$, 0.35% $H_3CO^+$, 0.25% $CO^+$, 0.25% $OH^+$, 0.2% $C_3H_3^+$, 0.7% $C_2H_2^+$, 0.23% $C_2H^+$, 0.13% $CH_3^+$, 0.13% $HCO^+$, 0.1% $S^+$, 0.05% $CH_3OH^+$, 0.05% $C_2^+$, 0.04% $O^+$, 0.04% $H_2S^+$, 0.04% $C_3H_4^+$, 0.04% $O_2^+$, 0.04% $C^+$, and 0.02% $CH_4^+$. Thus, the 24 important ions contribute 99% of the ion density in the inner coma of comet Halley, while only less than 1% is contributed by other 22 ions.

The calculated mass densities are compared with Giotto IMS and NMS data at radial distance 1500 km, 3500 km, and 6000 km. The nine major peaks at 12, 15, 19, 26, 28, 31, 33, 35 and 39 amu observed in the IMS spectra are produced mainly by ions $C^+$, $CH_3^+$, $H_3O^+$, $C_2H_2^+$, $H_2CN^+$, $H_3CO^+$, $CH_3OH_2^+$, $H_3S^+$, and $C_3H_3^+$, respectively. These peaks are reproduced well by model calculations inside the ionopause, except at masses 36–37 amu. The production and loss rates of these ions involve close to 180 chemical reactions.

There is a good agreement between the calculated electron density and those inferred from the Giotto measurement. It should be noted that all in-situ measurements on comet Halley were carried out at distances >1000 km from the nucleus. Therefore, we do not have experimental data near the cometary nucleus. Above 1000 km, the electron density is inferred from total ion density summed over all ions for mass range 12-56 amu.

To study the composition and electron density near the nucleus, we require a rendezvous mission, i.e. a nucleus orbiter. Only an orbiter provides the ability to map the entire nucleus surface from low cometocentric distance with high resolution.

The important and new aspects of the present study are:

1. We have explicitly evaluated the role of both: the photoelectron and the auroral electron impact ionization, relative to solar EUV, in the chemistry of the inner coma of the comet Halley. As described in text in Section 4 (cf. also Tables 2 to 6), the photoelectron ionization source plays an important role (and in fact exceeds the photoionization source in the region close to the nucleus (cf. also Bhardwaj, 2003)) in creating seed-ions, by direct and dissociative ionization of parent neutral species, that initiate the ion-neutral chemistry in the coma.

2. We have shown that the processes controlling the chemistry of several of the important ions are varying drastically as a function of the cometocentric distance in the inner coma. The major processes are quite different in the region below ~1000±500 km compared to those above this distance for most of the ions. For example, the electron dissociative recombination is an important loss mechanism, but it is generally the dominant loss process only at >~1000 km for most of the major ionic species in the coma. In particular, the new aspect of the present study is the presentation of the chemistry of the innermost (<1000 km) part of the coma of comet Halley.

3. We have reported simple formulae that can be used to calculate the density of the nine important ions, viz., $H_3O^+$, $NH_4^+$, $C_2H_2^+$, $CH_3OH_2^+$, $H_3S^+$, $H_2CN^+$, $H_2O^+$, $H_3CO^+$, and $C_3H_3^+$, approximately in the coma of comet Halley without running present chemical model of complex nature. These formulae are given in the text by equations (A1) to (A9). In Figure 1a we present the comparison between the density profiles of these ions obtained from the extensive chemical model with those obtained by using equations (A1) to (A9). The comparison suggests that the density profiles calculated by using (A1–A9) are a good representative of those obtained by running the comprehensive chemical model.

Earlier Wegmann et al. (1987) (cf. also Schmidt et al., 1988) have calculated the mass density of ions corresponding to ≤55 amu in the coma of comet Halley. However, they did not include the chemistry of $CH_3OH$, $H_2S$, and water dimer in their chemical model. We have found that a few percent of water dimer in the model increases sharply the density at masses 36 and 37 amu approaching close to IMS measurements. The density of other ion masses were found to be unaffected by including the dimer chemistry in the model. The methanol and hydrogen disulfide have been measured in several comets. These molecules play an important role in the chemistry of inner coma at larger ion masses as revealed from the present model calculations.

We now compare the chemistry of some of the major ions reported in this paper with those given in earlier work, where major production and loss processes are explicitly reported along with their relative contributions. We will mainly highlight the differences, if any, between the results of the current model and the earlier work.

### 5.1. $H_3O^+$ ion

Our model results are in general agreement with those of Eberhardt and Krankowsky (1995). However, our model show that losses of ion $H_3O^+$ in reaction with $H_2S$ and HCN are also significant — amounting to ~11% at 100 km and 4% at 1000 km. Also, the production of $H_3O^+$ ion due to the reaction of $C_2H_2^+$ with $H_2O$ makes a contribution of ~ 3% at 100 km and ~1% at 1000–5000 km, while $H_2CO^+ + H_2O$ reaction contribute ~1% at distances 1000–5000 km.

### 5.2. $NH_4^+$ ion

Though the model results are in qualitative agreement with those of Meier *et al*. (1994), there are differences between the two. Our model suggest a contribution of reaction $NH_3^+ + H_2O$ to the production of $NH_4^+$ ion of <1% throughout the inner coma, while Meier *et al*. (1994) reported its contribution to be 5.3% at 2500 km. Our model shows that to the production of $NH_4^+$ ion: 1) the reaction of $H_3S^+$ with $NH_3$ makes an important contribution of 12% at 100 km, which reduces to 3% at 1000 km and 1% at 5000 km, 2) the reaction of $H_2CN^+$ with $NH_3$ contributes 8% at 100 km and ~1-2% at 1000-2000 km, and 3) the reaction $H_3O^+ + NH_3$ contributes 1% at 100 km, which rises to 3% at 1000 km and 7% at 5000 km.

### 5.3. $H_3CO^+$ ion

The major production and loss mechanisms are in agreement with Meier *et al*. (1993). Two new production sources of $H_3CO^+$ ion revealed by the present study (cf. Figure 5) are reactions $CO^+ + CH_3OH$ and $CH_3^+ + CH_3OH$, which are the main loss process below 50 km, and contribute ~35% at 100 km. The first reaction also makes a significant contribution of about 3% and 8%, respectively, at 1000 and 5000 km. The contribution of reaction $CH_2^+ + H_2O$ to the production of $H_3CO^+$ ion is very low (<1%) compare to that (~14% at 2500 km) given by the model of Meier *et al*. (1993). Our model suggest a significant loss of $H_3CO^+$ ion in reaction with $NH_3$, amounting to about 10% at 100 km, 6% at 1000 km, and 2% at 5000 km.

### 5.4. $CH_3OH_2^+$ ion

The major production and loss channels are in agreement with the work of Eberhardt *et al*. (1994) (cf. also Eberhardt and Krankowsky, 1995). But, our model suggests that contribution from reactions $C_2H_2^+ + CH_3OH$ and $C_3H_4^+ + CH_3OH$ to the production of $CH_3OH_2^+$ ion is roughly equal to that from $HCO^+ + CH_3OH$ reaction (cf. Figure 6). Also, the current model shows that at distances <100 km the main loss of $CH_3OH_2^+$ ion is in reaction with $NH_3$ and $CH_3OH$.

### 5.5. $H_3S^+$ ion

The model results of Eberhardt *et al*. (1994) and ours are in agreement. But we have found that (cf. Figure 7) at distances close to the nucleus (<200 km) about 15% of $H_3S^+$ ion is lost in reactions with $H_2O$ and HCN, while the main loss occurs in reaction with $NH_3$. Similarly, for the production of $H_3S^+$ ion, at <500 km, the contribution from $H_2CN^+ + H_2S$ is also significant and in fact is larger than that due to the $H_2O^+ + H_2S$ reaction.

**Acknowledgements**

**Table 1. Ion masses corresponding to ≤40 amu**

| Mass | Ions |
|------|------|
| 12 | $C^+$ |
| 13 | $CH^+, {}^{13}C^+$ |
| 14 | $N^+, CH_2^+, {}^{13}CH^+$ |
| 15 | $NH^+, CH_3^+, {}^{13}CH_2^+$ |
| 16 | $NH_2^+, O^+, CH_4^+, {}^{13}CH_3^+$ |
| 17 | $OH^+, CH_5^+, NH_3^+, {}^{13}CH_4^+$ |
| 18 | $H_2O^+, NH_4^+$ |
| 19 | $H_3O^+$ |
| 20 | $H_2{}^{18}O^+$ |
| 21 | $H_3{}^{18}O^+$ |
| 22 | |
| 23 | |
| 24 | $C_2^+$ |
| 25 | $C_2H^+$ |
| 26 | $C_2H_2^+$ |
| 27 | $HCN^+, C_2H_3^+$ |
| 28 | $N_2^+, C_2H_4^+, CO^+, H_2CN^+$ |
| 29 | $N_2H^+, HCO^+, C_2H_5^+$ |
| 30 | $H_2CO^+, CH_2NH_2^+$ |
| 31 | $CH_3O^+, HNO^+$ |
| 32 | $O_2^+, S^+, CH_3OH^+$ |
| 33 | $SH^+, CH_3OH_2^+$ |
| 34 | $H_2S^+, {}^{34}S^+, {}^{13}CH_3OH_2^+$ |
| 35 | $H_3S^+$ |
| 36 | $H_2{}^{34}S^+, H_3{}^{33}S^+$ |
| 37 | $C_3H^+$ |
| 38 | $C_2N^+$ |
| 39 | $C_3H_3^+$ |
| 40 | $C_3H_4^+, CH_2CN^+$ |

**Table 2.** Major production and loss channels of $C^+$, $CH^+$, $CH_2^+$, $CH_3^+$, $CH_4^+$, and $CH_5^+$ ions and their relative contributions (in percentage) at 5 radial distances from the cometary nucleus.

| Production/Loss Reaction | Rate Constant (cm³/sec) | 100 (km) | 600 (km) | 1000 (km) | 2000 (km) | 5000 (km) |
|---|---|---|---|---|---|---|
| $C^+$ | | | | | | |
| $C^+ + H_2O$ | 2.54E-09 | 92.70 | 92.40 | 92.20 | 91.90 | 92.60 |
| $CO + h\nu$ | see text | 30.80 | 49.70 | 50.30 | 50.60 | 49.80 |
| $CO + PE$ | see text | 2.67 | 1.94 | 1.84 | 1.72 | 1.68 |
| $CO + AE$ | see text | 51.80 | 32.00 | 32.70 | 36.00 | 39.70 |
| $CO_2 + h\nu$ | | 4.60 | 7.42 | 6.91 | 5.33 | 4.01 |
| $CO_2 + PE$ | | <1 | <1 | <1 | <1 | <1 |
| $CO_2 + AE$ | | 1.53 | <1 | <1 | <1 | <1 |
| $CH^+$ | | | | | | |
| $CH^+ + H_2O$ | 2.90E-09 | 93.00 | 92.60 | 92.20 | 91.50 | 90.90 |
| $CH_4 + AE$ | see text | 12.20 | 8.44 | 8.83 | 9.65 | 10.50 |
| $C_2H_2 + PE$ | see text | 28.80 | 27.40 | 26.30 | 24.30 | 22.90 |
| $C_2H_2 + AE$ | see text | 29.90 | 20.00 | 20.90 | 23.30 | 26.00 |
| $C_2H_2 + h\nu$ | see text | 18.30 | 31.50 | 31.60 | 30.90 | 29.20 |
| $CH_2^+$ | | | | | | |
| $CH_2^+ + H_2O$ | 1.20E-09 | 85.00 | 83.50 | 81.70 | 78.90 | 75.20 |
| $C^+ + H_2CO$ | 2.34E-09 | <1 | 8.54 | 14.90 | 26.50 | 27.10 |
| $CH_4 + h\nu$ | see text | 17.50 | 28.30 | 26.40 | 22.30 | 20.70 |
| $CH_4 + PE$ | see text | 40.70 | 36.10 | 32.40 | 25.70 | 23.90 |
| $CH_4 + AE$ | see text | 41.30 | 26.60 | 26.10 | 25.40 | 28.20 |
| $CH_3^+$ | | | | | | |
| $CH_3^+ + C_2H_2$ | 1.20E-09 | 27.60 | 20.80 | 17.00 | 12.20 | 8.17 |
| $CH_3^+ + NH_3$ | 1.80E-09 | 24.50 | 17.00 | 13.00 | 7.92 | 3.22 |
| $CH_3^+ + CH_3OH$ | 2.30E-09 | 36.00 | 27.20 | 22.20 | 16.00 | 10.70 |
| $CH_3^+ + N_e$ | 3.50E-07*$(300/T_e)^{0.5}$ | 1.91 | 24.30 | 37.10 | 54.10 | 72.30 |
| $CH_4 + h\nu$ | see text | 16.50 | 28.10 | 28.00 | 27.10 | 25.00 |
| $CH_4 + PE$ | see text | 42.20 | 40.50 | 38.70 | 35.30 | 32.40 |
| $CH_4 + AE$ | see text | 32.40 | 22.30 | 24.00 | 27.30 | 30.90 |
| $CH_4^+$ | | | | | | |
| $CH_4^+ + H_2O$ | 2.50E-09 | 87.20 | 85.70 | 83.90 | 79.60 | 73.00 |
| $CH_4^+ + N_e$ | 6.00E-07*$(300/T_e)^{0.5}$ | <1 | 2.05 | 3.75 | 7.22 | 13.20 |
| $CH_4 + h\nu$ | see text | 18.50 | 30.30 | 29.90 | 28.80 | 26.40 |
| $CH_4 + PE$ | see text | 41.20 | 37.90 | 36.70 | 33.30 | 30.60 |
| $CH_4 + AE$ | see text | 27.80 | 18.70 | 20.20 | 23.60 | 27.50 |
| $CH_5^+$ | | | | | | |
| $CH_5^+ + H_2O$ | 3.70E-09 | 93.00 | 91.20 | 89.30 | 84.90 | 78.70 |
| $CH_5^+ + CO$ | 9.90E-10 | 4.99 | 4.92 | 5.26 | 6.58 | 8.22 |
| $CH_4^+ + CH_4$ | 1.50E-09 | 22.90 | 22.40 | 22.10 | 21.90 | 21.90 |
| $OH^+ + CH_4$ | 1.95E-10 | 74.00 | 74.80 | 75.40 | 75.90 | 76.70 |

Here $N_e$ = electron density, $h\nu$ = solar photon, PE = photoelectron, and AE = auroral electron, and 2.54E-09 = 2.54 x 10⁹.

**Table 3.** Major production and loss channels of $N^+$, $NH^+$, $NH_2^+$, $NH_3^+$, and $NH_4^+$ ions and their relative contributions (in percentage) at 5 radial distances from the cometary nucleus.

| Production/Loss Reaction | Rate Constant ($cm^3$/sec) | 100 (km) | 600 (km) | 1000 (km) | 2000 (km) | 5000 (km) |
|---|---|---|---|---|---|---|
| $N^+$ | | | | | | |
| $N^+ + H_2O$ | 2.80E-09 | 92.60 | 91.00 | 89.40 | 85.80 | 80.50 |
| $N_2 + h\nu$ | see text | <1 | 9.48 | 9.92 | 10.50 | 11.70 |
| $NH_3 + h\nu$ | see text | 16.00 | 27.70 | 27.00 | 24.00 | 15.70 |
| $NH_3 + AE$ | see text | 17.80 | 11.50 | 11.00 | 9.96 | 6.81 |
| $N_2 + PE$ | see text | 24.70 | 25.60 | 25.60 | 25.50 | 28.40 |
| $N_2 + AE$ | see text | 33.40 | 23.60 | 24.60 | 28.50 | 36.40 |
| $NH^+$ | | | | | | |
| $NH^+ + H_2O$ | 3.50E-09 | 91.50 | 91.00 | 90.00 | 85.50 | 86.50 |
| $NH_3 + h\nu$ | see text | 29.30 | 52.20 | 52.90 | 52.90 | 51.20 |
| $NH_3 + PE$ | see text | 15.70 | 12.50 | 11.90 | 11.00 | 10.00 |
| $NH_3 + AE$ | see text | 55.00 | 35.30 | 35.20 | 36.50 | 38.20 |
| $NH_2^+$ | | | | | | |
| $NH_2^+ + H_2O$ | 5.10E-09 | 96.60 | 95.50 | 94.60 | 92.70 | 89.80 |
| $NH_2^+ + N_e$ | $3.0E-07*(300/T_e)^{0.5}$ | <1 | 1.02 | 1.88 | 3.77 | 7.2 |
| $NH_3 + h\nu$ | see text | <1 | <1 | <1 | <1 | <1 |
| $NH_3 + PE$ | see text | 44.90 | 53.70 | 51.90 | 47.70 | 43.90 |
| $NH_3 + AE$ | see text | 53.80 | 44.20 | 46.10 | 50.00 | 54.30 |
| $NH_3^+$ | | | | | | |
| $NH_3^+ + H_2O$ | 2.50E-10 | 73.70 | 44.30 | 32.40 | 20.30 | 11.80 |
| $NH_3^+ + CH_3OH$ | 2.20E-09 | 11.00 | 6.62 | 4.84 | 3.04 | 1.76 |
| $NH_3^+ + N_e$ | $4.15E-05*(T_e)^{-0.5}$ | 4.19 | 42.40 | 57.80 | 73.60 | 85.20 |
| $C_2H_2^+ + NH_3$ | 2.14E-09 | 12.20 | 10.20 | 9.31 | 7.83 | 6.09 |
| $H_2O^+ + NH_3$ | 2.21E-09 | 34.50 | 35.60 | 36.20 | 37.20 | 38.40 |
| $CO^+ + NH_3$ | 1.99E-09 | 4.92 | 4.67 | 5.15 | 6.90 | 9.61 |
| $NH3 + h\nu$ | see text | 15.00 | 22.90 | 23.40 | 23.70 | 23.60 |
| $NH_3 + PE$ | see text | 8.71 | 7.41 | 7.23 | 6.89 | 6.87 |
| $NH_3 + AE$ | see text | 8.26 | 4.86 | 5.18 | 5.99 | 7.10 |
| $NH_4^+$ | | | | | | |
| $NH_4^+ + N_e$ | $4.15E-05*(T_e)^{-0.5}$ | 100.0 | 100.0 | 100.0 | 100.0 | 100.0 |
| $NH_3^+ + H_2O$ | 2.50E-10 | 5.03 | 5.70 | 5.76 | 5.60 | 5.23 |
| $H_3O^+ + NH_3$ | 2.20E-9 | 70.80 | 78.80 | 79.90 | 79.70 | 79.00 |
| $CH_3O^+ + NH_3$ | 1.27E-09 | <1 | 1.91 | 3.36 | 5.39 | 4.41 |
| $H_3S^+ + NH_3$ | 1.90E-09 | 12.90 | 5.64 | 3.44 | 1.54 | <1 |
| $H_2O^+ + NH_3$ | 9.0E-10 | 1.01 | 1.97 | 2.77 | 4.40 | 7.28 |
| $H_2CN^+ + NH_3$ | 2.4E-09 | 7.92 | 3.41 | 2.24 | 1.29 | <1 |
| $CH_3O^+ + NH_3$ | 1.27E-09 | <1 | 1.6 | 2.48 | 3.15 | 1.79 |

Here $N_e$ = electron density, $h\nu$ = solar photon, PE = photoelectron, and AE = auroral electron, and 2.80E-09 = 2.80 x $10^9$.

**Table 4.** Major production and loss channels of $O^+$, $OH^+$, $H_2O^+$, and $H_3O^+$ ions and their relative contributions (in percentage) at 5 radial distances from the cometary nucleus.

| Production/Loss Reaction | Rate Constant ($cm^3$/sec) | 100 (km) | 600 (km) | 1000 (km) | 2000 (km) | 5000 (km) |
|---|---|---|---|---|---|---|
| \multicolumn{7}{c}{$O^+$} | | | | | | |
| $O^+ + H_2O$ | 2.50E-09 | 96.2 | 95.9 | 95.7 | 95.4 | 95.9 |
| $O + h\nu$ | see text | 8.66 | 16.1 | 16.1 | 14.4 | 12.4 |
| $H_2O + AE$ | see text | 47.5 | 32.5 | 31.1 | 28.0 | 24.6 |
| $H_2O + PE$ | see text | 5.8 | 4.53 | 4.18 | 3.52 | 2.98 |
| $H_2O + h\nu$ | see text | <1 | <1 | <1 | <1 | <1 |
| $CO + h\nu$ | see text | 10.90 | 20.30 | 21.90 | 26.20 | 30.20 |
| $CO + PE$ | see text | 12.3 | 11.3 | 11.6 | 12.9 | 14.8 |
| $CO + AE$ | see text | 6.19 | 4.16 | 4.29 | 5.09 | 6.10 |
| $CO_2 + h\nu$ | see text | 4.32 | 8.01 | 7.95 | 7.30 | 6.45 |
| $CO_2 + AE$ | see text | 3.09 | 1.99 | 1.88 | 1.75 | 1.66 |
| $CO_2 + PE$ | see text | 1.06 | 1.01 | <1 | <1 | <1 |
| \multicolumn{7}{c}{$OH^+$} | | | | | | |
| $OH^+ + H_2O$ | 2.89E-09 | 93.20 | 93.00 | 92.40 | 90.70 | 88.60 |
| $OH^+ + CO$ | 7.1E-10 | 4.59 | 4.6 | 6.0 | 6.45 | 8.49 |
| $OH^+ + CO_2$ | 1.1E-09 | 1.07 | 1.07 | 1.07 | 1.06 | 1.05 |
| $H_2O + h\nu$ | see text | 29.70 | 47.90 | 48.50 | 47.20 | 45.00 |
| $H_2O + PE$ | see text | 31.20 | 26.30 | 24.50 | 22.50 | 21.40 |
| $H_2O + AE$ | see text | 39.1 | 25.80 | 26.90 | 30.20 | 33.50 |
| \multicolumn{7}{c}{$H_2O^+$} | | | | | | |
| $H_2O^+ + H_2O$ | 2.05E-09 | 93.80 | 92.50 | 91.00 | 87.70 | 82.90 |
| $H_2O^+ + NH_3$ | 3.15E-09 | 2.13 | 1.93 | 1.78 | 1.45 | <1 |
| $H_2O^+ + CO$ | 3.6E-10 | 3.30 | 3.27 | 3.52 | 4.46 | 5.68 |
| $H_2O^+ + N_e$ | $3.15E-07*(300/T_e)^{0.5}$ | <1 | 1.42 | 2.61 | 5.09 | 9.58 |
| $H_2O + h\nu$ | see text | 30.00 | 43.80 | 43.30 | 41.10 | 37.50 |
| $H_2O + PE$ | see text | 16.50 | 13.40 | 12.70 | 11.30 | 10.30 |
| $H_2O + AE$ | see text | 16.20 | 9.97 | 10.70 | 11.90 | 13.20 |
| $OH^+ + H_2O$ | 1.58E-09 | 9.68 | 8.77 | 8.78 | 8.38 | 7.83 |
| $CO^+ + H_2O$ | 1.69E-09 | 12.20 | 11.00 | 11.80 | 14.80 | 18.90 |
| $O^+ + H_2O$ | 2.5E-09 | 3.07 | 2.4 | 2.41 | 2.51 | 2.69 |
| $H^+ + H_2O$ | 8.2E-09 | 8.74 | 7.27 | 7.22 | 6.84 | 6.59 |
| \multicolumn{7}{c}{$H_3O^+$} | | | | | | |
| $H_3O^+ + NH_3$ | 2.20E-09 | 36.00 | 18.40 | 12.20 | 6.43 | 2.34 |
| $H_3O^+ + CH_3OH$ | 2.50E-09 | 47.00 | 26.20 | 18.60 | 11.50 | 6.92 |
| $H_3O^+ + N_e$ | $6.99E-07*(300/T_e)^{0.5}$ | 4.60 | 43.00 | 57.10 | 71.80 | 85.90 |
| $H_3O^+ + H_2S$ | 1.65E-09 | 7.29 | 3.49 | 2.20 | 1.01 | <1 |
| $H_3O^+ + HCN$ | 4.5E-09 | 4.21 | 2.31 | 1.66 | 1.03 | <1 |
| $H_3O^+ + H_2CO$ | 3.4E-09 | <1 | 6.6 | 8.19 | 8.24 | 4.01 |
| $H_2O^+ + H_2O$ | 2.05E-09 | 75.60 | 73.60 | 72.70 | 71.70 | 71.90 |
| $HCO^++H_2O$ | 3.20E-09 | 9.41 | 8.67 | 9.31 | 11.90 | 15.60 |
| $OH^++H_2O$ | 2.89E-09 | 6.38 | 5.72 | 5.74 | 5.61 | 5.56 |
| $C_2H_2^++H_2O$ | 2.20E-10 | 2.97 | 2.34 | 2.07 | 1.67 | 1.26 |
| $CH_3O^++H_2O$ | 2.30E-10 | <1 | 4.98 | 5.4 | 4.27 | 1.47 |

Here $N_e$ = electron density, hν = solar photon, PE = photoelectron, and AE = auroral electron, and 2.50E-09 = 2.50 x $10^9$.

**Table 5.** Major production and loss channels of $C_2^+$, $C_2H^+$, $CO^+$, $HCO^+$, $CH_3OH^+$, and $C_3H_4^+$ ions and their relative contributions (in percentage) at 5 radial distances from the cometary nucleus.

| Production/Loss Reaction | Rate Constant (cm$^3$/sec) | 100 (km) | 600 (km) | 1000 (km) | 2000 (km) | 5000 (km) |
|---|---|---|---|---|---|---|
| colspan="7" | | | | | | |
| $C_2^+$ | | | | | | |
| $C_2^+ + C_2H_2$ | 1.20E-09 | 80.90 | 32.40 | 20.90 | 11.70 | 6.24 |
| $C_2^+ + N_e$ | 5.90E-07*(300/$T_e$)$^{0.5}$ | 9.44 | 63.70 | 76.60 | 86.20 | 93.0 |
| $C_2H_2$+PE | see text | 49.00 | 57.80 | 55.70 | 51.00 | 46.70 |
| $C_2H_2$+AE | see text | 51.00 | 42.20 | 44.30 | 49.00 | 53.20 |
| $C_2H^+$ | | | | | | |
| $C_2H^+ + C_2H_2$ | 2.40E-09 | 86.00 | 63.00 | 50.60 | 35.20 | 22.00 |
| $C_2H^+ + N_e$ | 2.70E-07*(300/$T_e$)$^{0.5}$ | 2.29 | 28.40 | 42.50 | 60.00 | 75.00 |
| $C_2H_2$+hν | see text | 23.80 | 39.90 | 40.10 | 39.40 | 37.40 |
| $C_2H_2$+PE | see text | 37.40 | 34.80 | 33.40 | 30.90 | 29.30 |
| $C_2H_2$+AE | see text | 38.80 | 25.30 | 26.50 | 29.70 | 33.30 |
| $CO^+$ | | | | | | |
| $CO^+ + H_2CO$ | 3.00E-09 | 3.84 | 59.10 | 80.10 | 94.20 | 99.00 |
| $CO^+ + H2O$ | 2.60E-09 | 91.50 | 38.90 | 18.90 | 5.49 | <1 |
| CO+hν | see text | 27.80 | 44.80 | 45.10 | 44.40 | 42.50 |
| CO+PE | see text | 34.00 | 30.80 | 29.60 | 27.50 | 26.20 |
| CO+AE | see text | 37.10 | 22.80 | 23.60 | 26.50 | 30.20 |
| $HCO^+$ | | | | | | |
| $HCO^+ + H_2O$ | 3.20E-09 | 96.30 | 95.20 | 94.30 | 92.80 | 91.80 |
| $HCO^+ + N_e$ | 2.00E-07*(300/$T_e$)$^{0.5}$ | <1 | 1.03 | 1.85 | 3.27 | 4.77 |
| $C^+ + H_2O$ | 2.30E-09 | 9.40 | 9.03 | 8.88 | 8.70 | 8.74 |
| $H_2O^+ + CO$ | 3.60E-10 | 27.20 | 28.60 | 28.50 | 28.50 | 29.00 |
| $CO^+ + H_2O$ | 9.10E-10 | 54.20 | 52.00 | 51.30 | 51.00 | 51.90 |
| $CH_3OH^+$ | | | | | | |
| $CH_3OH^+ + H_2O$ | 1.50E-09 | 99.80 | 96.80 | 94.10 | 88.80 | 80.00 |
| $CH_3OH^+ + N_e$ | 5.0E-07*(300/$T_e$)$^{0.5}$ | <1 | 3.22 | 5.85 | 11.2 | 20.0 |
| $CH_3OH$+PE | see text | 30.50 | 27.40 | 26.60 | 24.30 | 22.80 |
| $CH_3OH$+AE | see text | 20.60 | 13.50 | 14.60 | 17.20 | 20.50 |
| $CH_3OH$+hν | see text | 24.80 | 39.60 | 39.10 | 37.90 | 35.70 |
| $C_3H_4^+$ | | | | | | |
| $C_3H_4^+ + C_2H_2$ | 4.90E-10 | 80.50 | 19.70 | 11.60 | 6.07 | 3.13 |
| $C_3H_4^+ + N_e$ | 5.00E-07*(300/$T_e$)$^{0.5}$ | 19.50 | 80.30 | 88.40 | 93.90 | 96.90 |
| $C_2H^+ + CH_4$ | 1.32E-10 | 25.20 | 21.30 | 19.40 | 17.00 | 14.80 |
| $C_2H_2^+ + CH_4$ | 1.87E-10 | 74.80 | 78.70 | 80.60 | 83.00 | 85.20 |

Here $N_e$ = electron density, hν = solar photon, PE = photoelectron, and AE = auroral electron, and 1.20E-09 = 1.20 x 10$^9$.

**Table 6.** Major production and loss channels of $S^+$, $SH^+$, and $H_2S^+$ ions and their relative contributions (in percentage) at 5 radial distances from the cometary nucleus.

| Production/Loss Reaction | Rate Constant ($cm^3$/sec) | 100 (km) | 600 (km) | 1000 (km) | 2000 (km) | 5000 (km) |
|---|---|---|---|---|---|---|
| $S^+$ | | | | | | |
| $S^+ + C_2H_2$ | 9.80E-10 | 46.70 | 46.90 | 47.10 | 48.40 | 58.80 |
| $S^+ + NH_3$ | 1.60E-09 | 45.00 | 41.70 | 39.10 | 34.10 | 25.20 |
| $SO_2 + h\nu$ | see text | 1.71 | 3.71 | 4.14 | 4.64 | 5.69 |
| $SO_2 + PE$ | see text | 5.53 | 6.54 | 6.98 | 7.36 | 8.99 |
| $SO_2 + AE$ | see text | 7.40 | 6.00 | 6.74 | 8.20 | 11.50 |
| $H_2S + PE$ | see text | 11.90 | 13.60 | 14.00 | 13.30 | 12.70 |
| $H_2S + h\nu$ | see text | 24.60 | 49.60 | 53.20 | 53.70 | 48.50 |
| $H_2S + AE$ | see text | 11.90 | 9.14 | 10.10 | 11.60 | 12.50 |
| $CS_2 + h\nu$ | see text | <1 | <1 | <1 | <1 | <1 |
| $CS_2 + PE$ | see text | 22.7 | 7.29 | 2.65 | 13.3 | <1 |
| $CS_2 + AE$ | see text | 13.7 | 3.18 | 1.33 | <1 | <1 |
| $SH^+$ | | | | | | |
| $SH^+ + H_2O$ | 7.80E-10 | 93.10 | 91.10 | 89.30 | 85.80 | 80.00 |
| $SH^+ + N_e$ | 2.0E-07*$(300/T_e)^{0.5}$ | <1 | 2.33 | 4.27 | 8.3 | 15.4 |
| $H_2S + h\nu$ | see text | 11.80 | 24.10 | 25.90 | 26.20 | 23.60 |
| $H_2S + PE$ | see text | 1.73 | 19.70 | 20.20 | 19.30 | 17.40 |
| $H_2S + AE$ | see text | 16.90 | 13.00 | 14.50 | 16.80 | 18.20 |
| $H_2S^+$ | | | | | | |
| $H_2S^+ + N_e$ | 3.00E-07*$(300/T_e)^{0.5}$ | <1 | 3.52 | 6.40 | 12.20 | 21.90 |
| $H_2S^+ + H_2O$ | 7.94E-10 | 96.20 | 93.40 | 90.80 | 85.60 | 77.00 |
| $C_2H_2^+ + H_2S$ | 2.21E-09 | 18.10 | 15.00 | 13.60 | 11.10 | 8.20 |
| $CO^+ + H_2S$ | 2.44E-09 | 8.65 | 8.17 | 8.91 | 11.60 | 15.30 |
| $H_2O^+ + H_2S$ | 8.10E-10 | 18.20 | 18.70 | 18.80 | 18.70 | 18.40 |
| $H_2S + h\nu$ | see text | 19.80 | 30.10 | 30.40 | 29.80 | 28.20 |
| $H_2S + PE$ | see text | 17.10 | 15.00 | 14.50 | 13.50 | 12.80 |
| $H_2S + AE$ | see text | 12.20 | 7.33 | 7.91 | 9.29 | 11.00 |

Here $N_e$ = electron density, $h\nu$ = solar photon, PE = photoelectron, and AE = auroral electron, and 9.80E-10 = 9.80 x $10^{10}$.

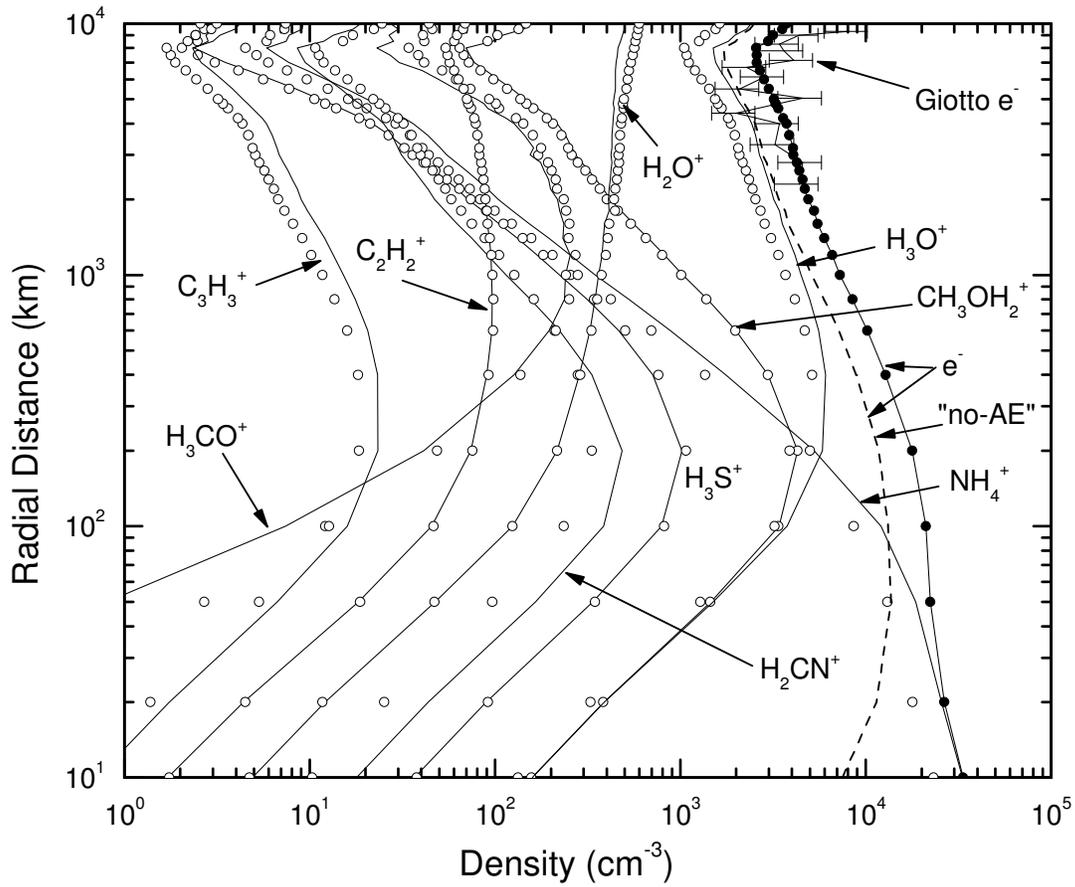

**Figure 1a.** The radial density profiles of 9 important ions $H_3O^+$, $NH_4^+$, $CH_3OH_2^+$, $H_3S^+$, $H_2CN^+$, $H_2O^+$, $C_3H_3^+$, $H_3CO^+$, and $C_2H_2^+$. The lines shown by open circles that closely follow the solid lines (marked for the respective ions) corresponds to the density profile for that ion obtained by using the equations (A1–A9) given in the text. The total ion density (equal to electron density) is shown for with and without auroral electron ("no-AE") cases. The Giotto measured electron density (with error bars) along the inbound pass is also shown in the figure.

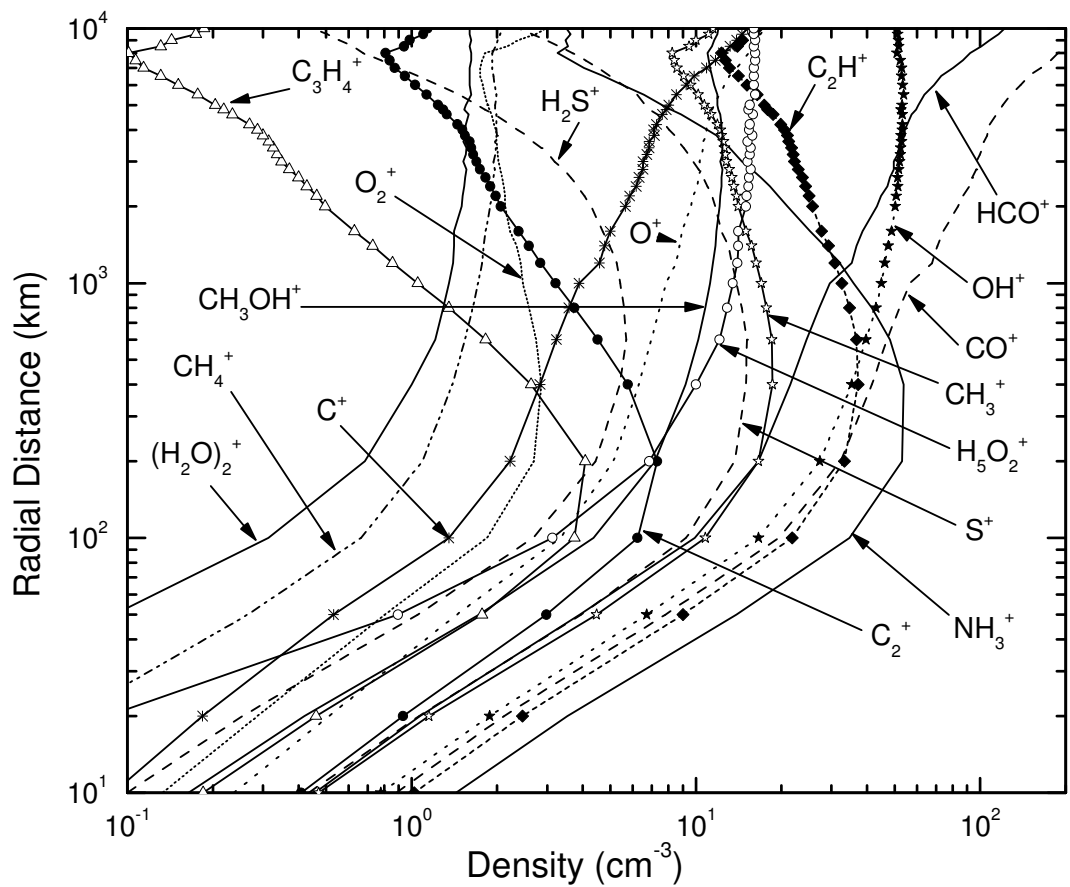

**Figure 1b.** The radial density profiles of 17 significant ions $CH_3OH^+$, $C_3H_4^+$, $C_2H^+$, $HCO^+$, $S^+$, $CH_3^+$, $H_2S^+$, $O^+$, $C^+$, $CH_4^+$, $C_2^+$, $O_2^+$, $NH_3^+$, $CO^+$, $OH^+$, $(H_2O)_2^+$, and $H_5O_2^+$. The densities of $(H_2O)_2^+$ and $H_5O_2^+$ are shown for 1% (relative to water) dimer case.

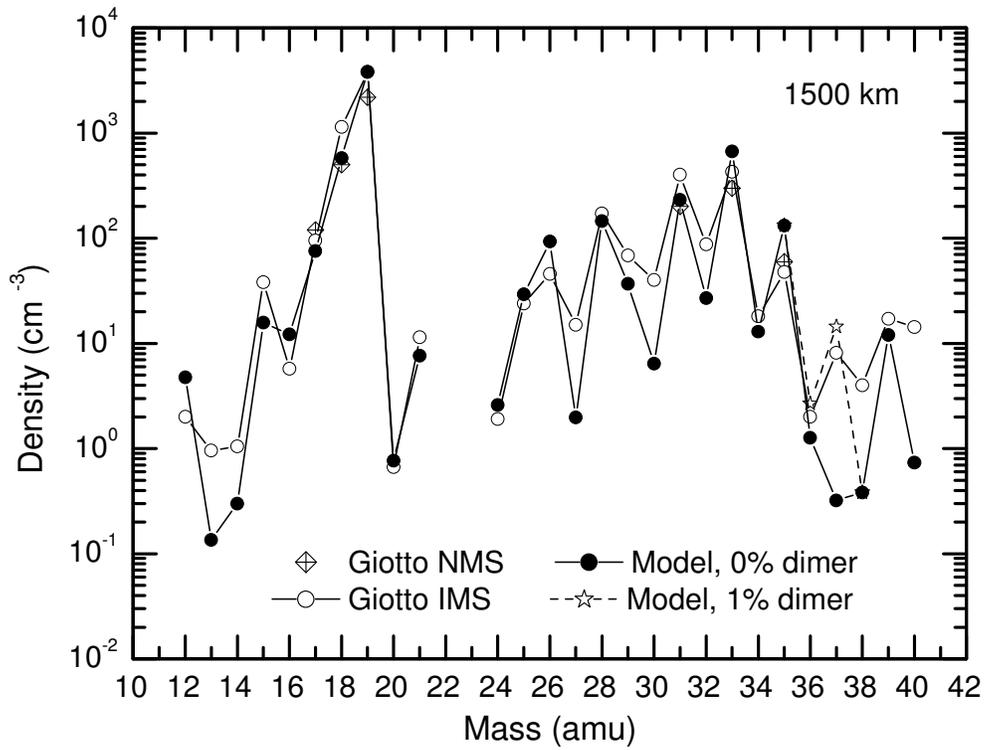

**Figure 2a.** Ion density spectra for masses 12 to 40 amu at radial distance 1500 km. The Giotto IMS and NMS data are also presented in this figure. The model calculations are presented for without (0%) and with (1%) dimer cases.

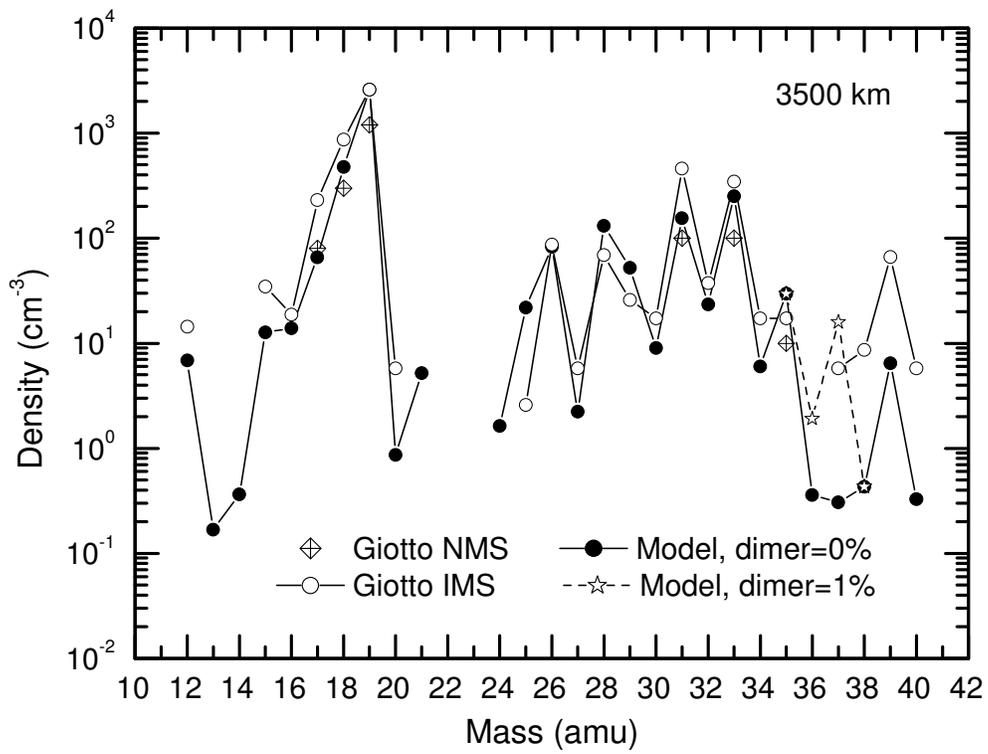

**Figure 2b.** Same as shown in the Figure 2a, but at radial distance 3500 km.

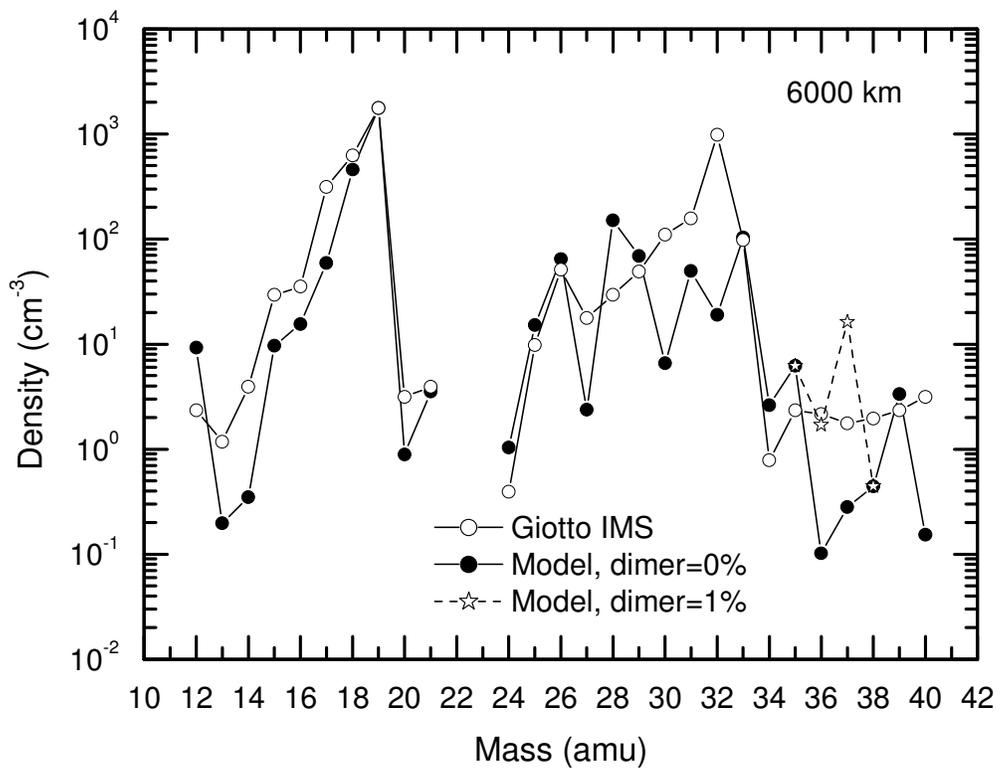

**Figure 2c.** Same as shown in the Figure 2a, but at radial distance 6000 km. In this figure, only the Giotto IMS data is shown.

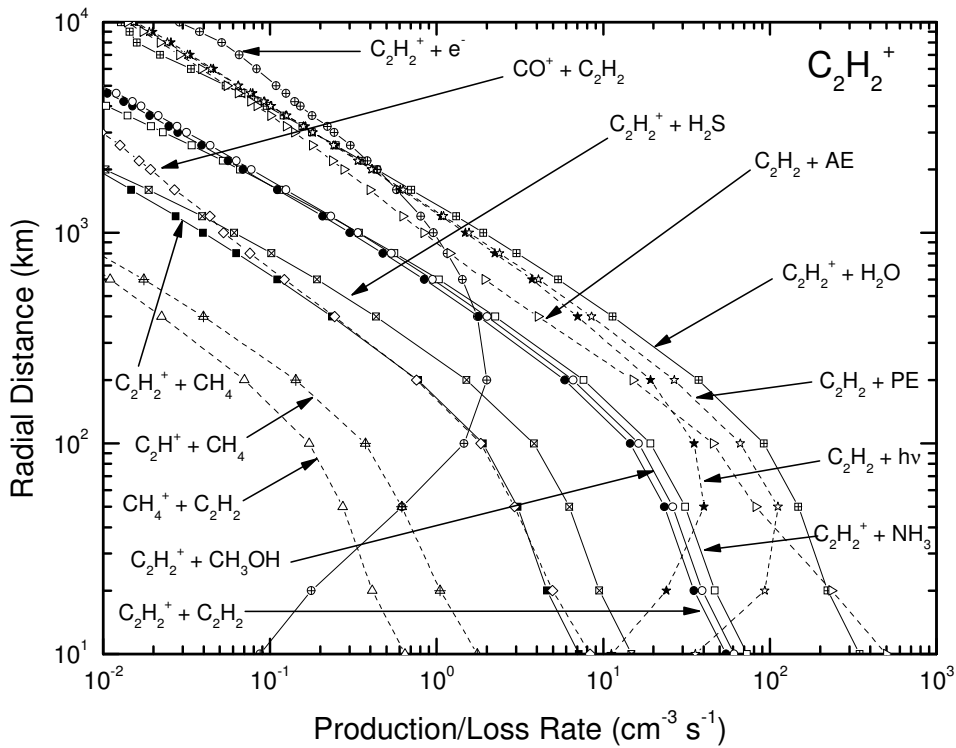

**Figure 3.** The production and loss rates of $C_2H_2^+$ as a function of radial distance for important chemical processes. Solid lines represent the loss reactions while the dashed lines show the production reactions.

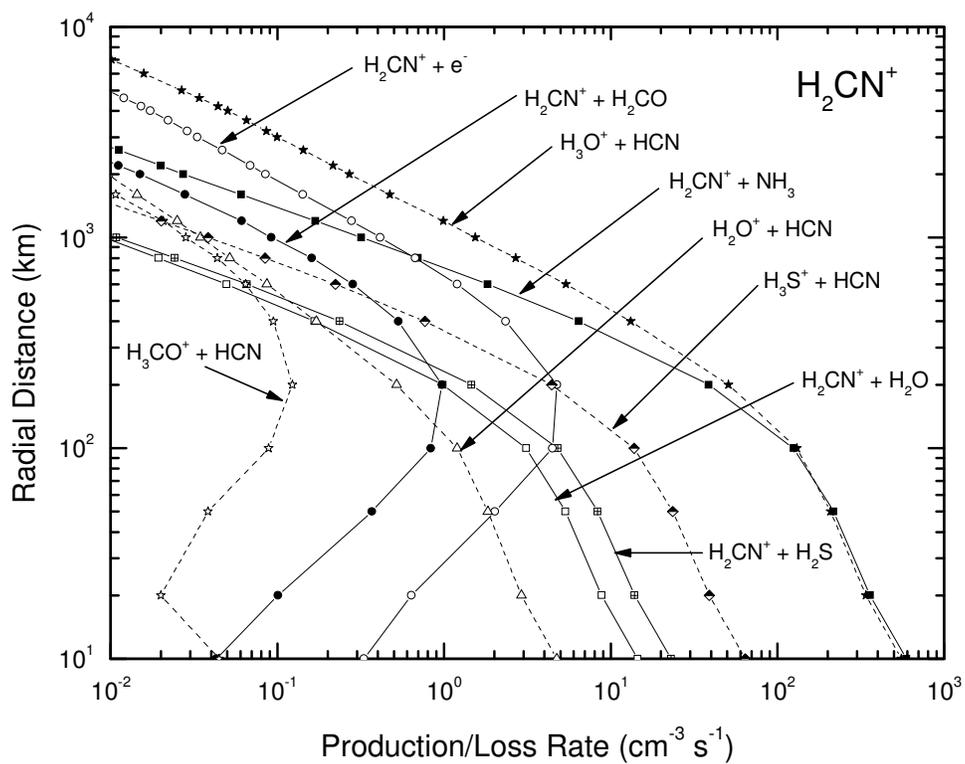

**Figure 4.** The production and loss rates of $H_2CN^+$ as a function of radial distance for major chemical processes in the chemistry of this ion. Solid lines represent the loss reactions while the dashed lines show the production reactions.

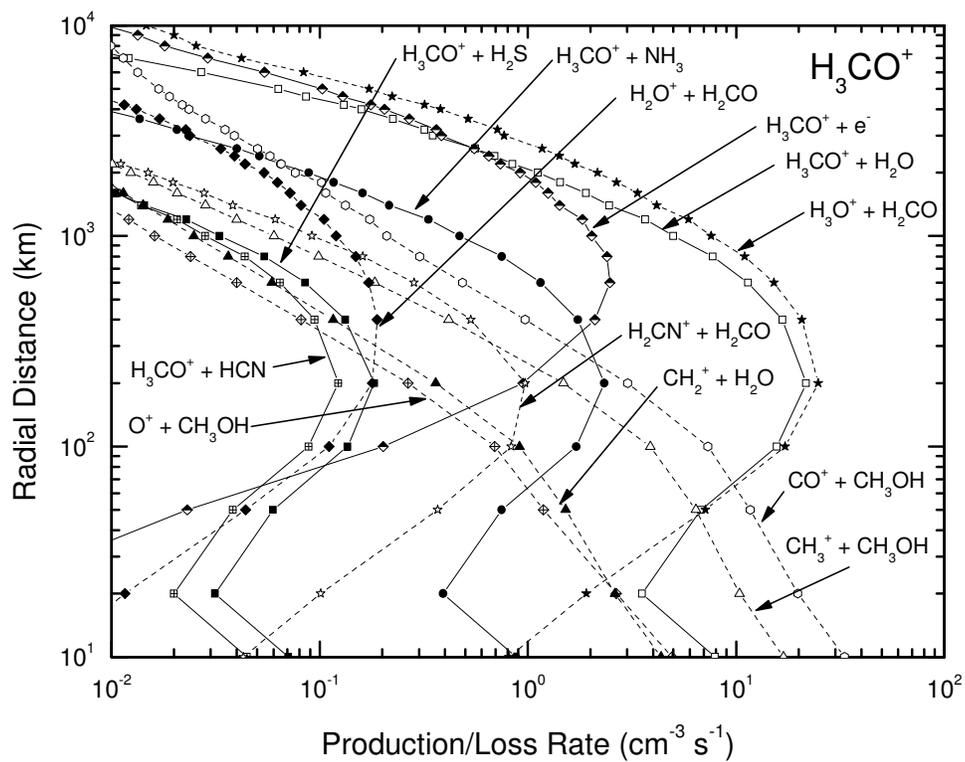

**Figure 5.** The production and loss rates of $H_3CO^+$ as a function of radial distance for important chemical reactions. Solid lines represent the loss reactions while the dashed lines show the production reactions.

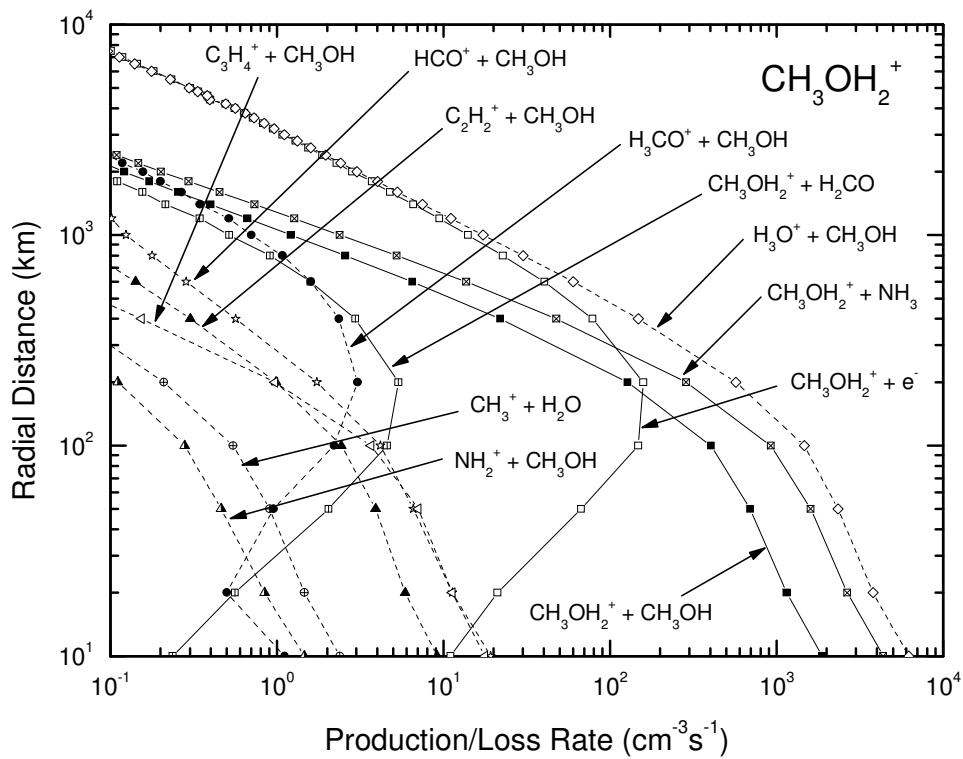

**Figure 6.** The production and loss rates of $CH_3OH_2^+$ as a function of radial distance for major chemical processes taking part in the chemistry of this ion. Solid lines represent the loss reactions while the dashed lines show the production reactions.

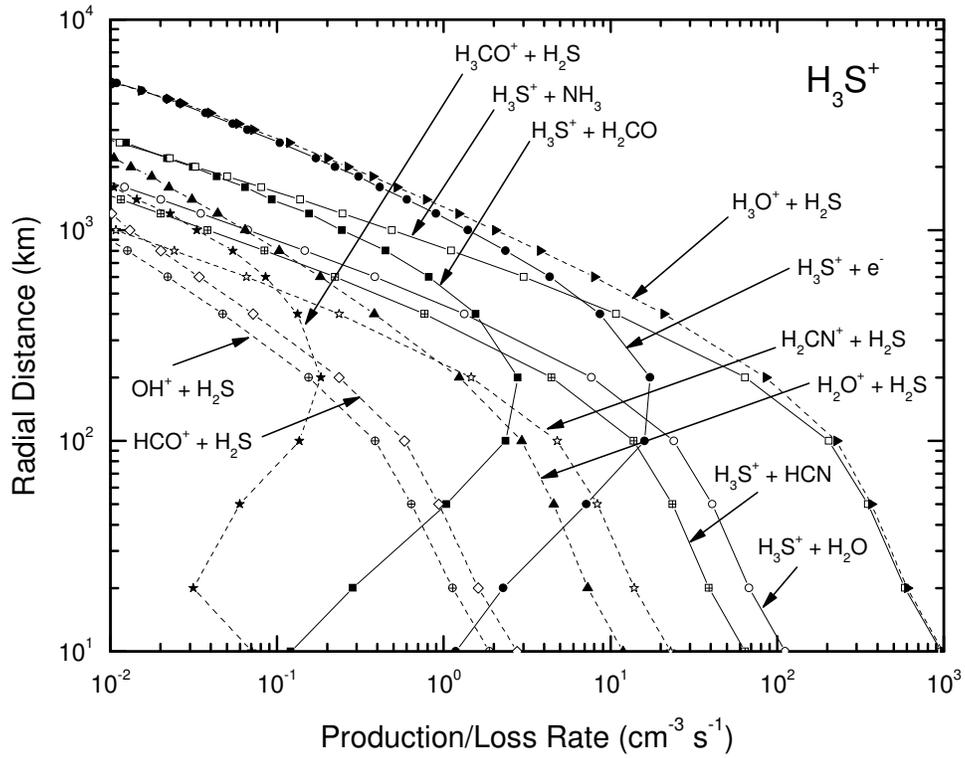

**Figure 7.** The production and loss rates of $H_3S^+$ as a function of radial distance for chemical processes that are important in the chemistry of this ion. Solid lines represent the loss reactions while the dashed lines show the production reactions.

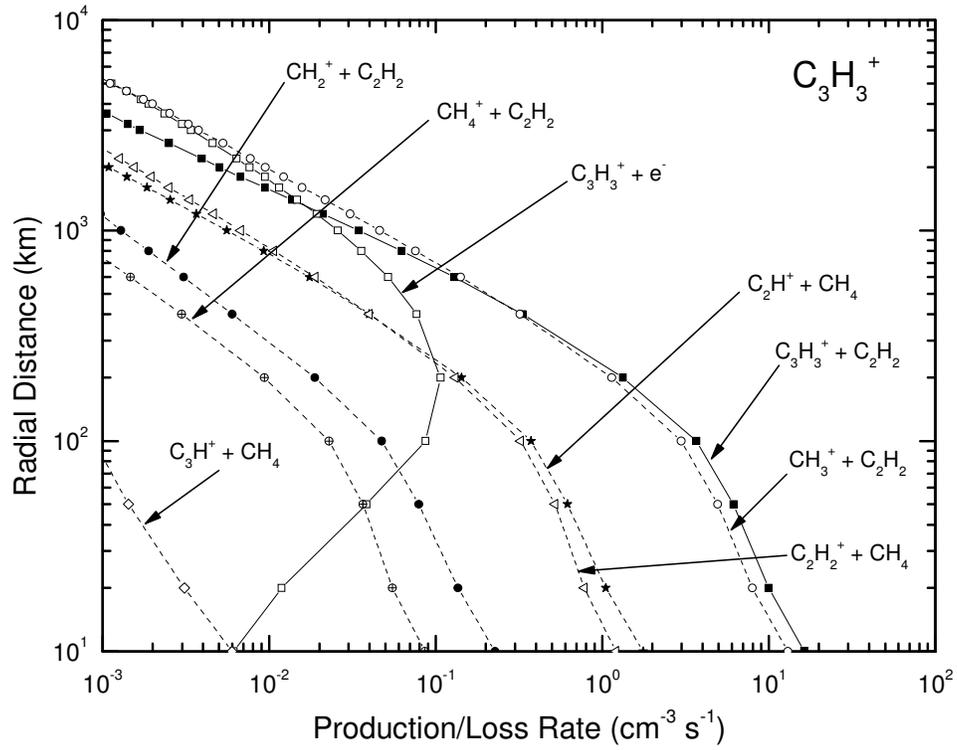

**Figure 8.** The production and loss rates of $C_3H_3^+$ as a function of radial distance presented for major chemical reactions. Solid lines represent the loss reactions while the dashed lines show the production reactions.

**Figure 9.** A schematic diagram illustrating the reaction pathways that are important in the chemistry of cometary coma. The primary sources of ionization, viz., solar EUV photon, photoelectron, and auroral electron that initiate the ion-neutral chemistry in the coma are given in the center within the circle. The symbol e⁻ refers to electrons. The ions, taking part in the coma chemistry, are outlined by rectangles.